\documentclass[aps,pra,10pt,twocolumn,groupaddress,longbibliography,nofootinbib,floatfix]{revtex4-2}

\usepackage[T1]{fontenc}
\usepackage{lmodern}
\usepackage[utf8]{inputenc}
\usepackage{amsmath,amssymb,amsfonts,bm}
\usepackage{graphicx}
\usepackage{xcolor}
\usepackage{hyperref}
\usepackage{braket}
\usepackage[percent]{overpic}

\definecolor{modelblue}{RGB}{0,70,130}
\definecolor{modelhot}{RGB}{195,72,31}
\definecolor{modelcold}{RGB}{25,110,185}
\definecolor{modelpurple}{RGB}{122,65,168}

\hypersetup{
  colorlinks=true,
  linkcolor=red,
  citecolor=blue,
  urlcolor=blue,
  pdftitle={Optimal phase control of a measurement-assisted quantum refrigerator},
  pdfauthor={Jinkai Liu, Xian He, Jizhou He, and Jianhui Wang},
  pdfsubject={Randomized open-loop phase cycling in a measurement-assisted quantum refrigerator}
}

\newcommand{\Tr}{\operatorname{Tr}}

\begin{document}

\title{Optimal phase control of a measurement-assisted quantum refrigerator}

\author{Jinkai Liu}
\author{Xian He}
\author{Jizhou He}
\author{Jianhui Wang}
\email{wangjianhui@ncu.edu.cn}
\affiliation{Department of Physics, Nanchang University, Nanchang 330031, China}

\date{\today}

\begin{abstract}
A tilted nonselective measurement can close the upper transition of a
three-level quantum refrigerator, but it also generates coherence that
feeds back on the populations and suppresses the cooling current.
We formulate randomized open-loop control of the measurement azimuthal
phase \(\alpha\), sampled independently at each event from a probability
distribution \(P(\alpha)\), while keeping the event rate, polar angle,
and single-event population-transfer probability fixed.
Direct population transfer is phase independent, whereas the coherence
source carries the factor \(e^{-i\alpha}\).
Exact elimination of the stationary coherence gives
\(0\leq k_{\rm eff}[P]\leq A_m\), where \(k_{\rm eff}[P]\) is the
effective population-transfer rate between the upper levels and
\(A_m\) is the bare direct mixing rate.
When the fixed measurement couples populations and coherence, the
upper bound is attained if and only if
\(\langle e^{i\alpha}\rangle_P=0\), where
\(\langle\cdot\rangle_P\) denotes an average over the phase distribution.
The minimal realization uses two equally weighted phases separated
by \(\pi\): the population-to-coherence source and its return to the
populations then vanish at the generator level, while the full direct
mixing remains.
Because the cold current increases strictly with \(k_{\rm eff}\) in
the refrigeration window, the same condition maximizes the stationary
cooling power over the independent random-phase distributions
considered here.
For one representative parameter set, phase cycling increases the
fixed-gap cooling power by about \(32\%\) relative to a fixed-phase
measurement, while leaving the ratio of extracted cold heat to
supplied measurement energy unchanged.
\end{abstract}

\maketitle

\section{Introduction}

Quantum thermal machines provide controlled settings for studying energy
conversion and open-system dynamics
\cite{Alicki1979,Kosloff2014,Mitchison2019}. Their information-theoretic
aspects have been reviewed in Refs.~\cite{Goold2016,Vinjanampathy2016}.
Cycle-based and continuous-operation descriptions clarify which features are
shared by different architectures~\cite{Quan2007,Uzdin2015}.
Three-level working media provide a simple setting in which two thermal
transitions exchange heat with cold and hot reservoirs, while a third process
closes the cycle~\cite{Scovil1959,Geusic1967,Palao2001}.
Related autonomous refrigerators have been analyzed through minimal models
and virtual temperatures~\cite{Linden2010,Levy2012,Brunner2012}; a stochastic
cycle description has also been developed for the three-level
machine~\cite{LiuGu2025}. Performance bounds~\cite{Correa2013,Clivaz2019}
and steady-state coherence effects~\cite{Correa2014,Kilgour2018} have been
studied alongside transient cooling~\cite{Brask2015,Mitchison2015}.
Experimental progress~\cite{Myers2022} includes refrigerators implemented
with trapped ions and superconducting circuits~\cite{Maslennikov2019,Aamir2025}.
The cooling power depends not only on the level spacings and bath rates, but
also on how the third process transfers population between the two upper
states.

Quantum measurement can provide this third process. In the standard
measurement framework~\cite{Jacobs2006,Wiseman2010}, a basis that does not
commute with the Hamiltonian can change the state and its mean energy even
when the outcome is discarded
\cite{Elouard2017NPJ,Kammerlander2016,Mohammady2021}.
This observation has motivated measurement-powered engines with two-level
and multilevel working media~\cite{Yi2017,Elouard2018PRL,Anka2021}, as well
as measurement-based cooling~\cite{Chand2017,Buffoni2019}.
Related protocols use entanglement~\cite{Bresque2021} or incompatible
measurements~\cite{Manikandan2022}. Outcome-conditioned feedback provides
another control route~\cite{Elouard2017PRL,Bhandari2023,LiuPRA2025}.
Information-powered refrigerator models offer a complementary route to
cooling without a conventional work drive~\cite{Mandal2013}.
Indefinite causal order provides a further control setting~\cite{Dieguez2023},
and experiments have demonstrated both generalized-measurement engines and
indefinite-order refrigerators~\cite{Lisboa2022,Nie2022}.

The role of coherence in thermal-machine performance is context dependent.
It can enhance work output~\cite{Scully2011,LuWang2024} and enable hybrid
operating regimes~\cite{Hammam2024}, but it can also generate power
losses~\cite{Brandner2017}. Its effects on finite-time Otto engines depend
on the driving protocol~\cite{Camati2019,Chen2019}.
Measurement-basis control has been used to suppress coherence-related
losses~\cite{Lin2021} and to optimize performance through the connection
between measurement and nonequilibrium free energy~\cite{Su2023}.
Here we consider a steady-state measurement-assisted refrigerator and vary
only the open-loop phase statistics. The polar angle, event rate, and
single-event population transfer are held fixed, so the control cannot
improve the current merely by strengthening the measurement.

A fixed tilted measurement combines two actions. The upper-state population
imbalance drives a direct population flow, but it also generates a coherence
whose direction is selected by the azimuthal phase. That coherence then feeds
back on the same population current. Increasing the measurement rate
strengthens both paths and therefore cannot separate them. The useful and
unwanted actions can instead be distinguished by their phase dependence: the
direct transition probability contains only squared projection amplitudes,
whereas the coherence source rotates with the azimuthal phase. We exploit this
difference by selecting the phase independently before each nonselective
measurement event. This randomized open-loop protocol is a form of phase
cycling. Phase cycling and selective averaging are established tools for
isolating coherence pathways~\cite{Ernst1987}. Related
symmetry-based and randomized-control methods suppress unwanted dynamics by
averaging over control operations
~\cite{Zanardi1999,ViolaLloyd1998,ViolaKnill2005,SantosViola2005}. The present
use is more specific:
the averaged operation retains the phase-independent transition probability of
\emph{each} measurement event, while eliminating both directions of a
population--coherence--population feedback loop. It is therefore not
equivalent to adding dephasing after the measurement or increasing the bare
transition rate.

We first define the full class of independent phase distributions
\(P(\alpha)\), then use a balanced two-phase family to expose the cancellation
mechanism, and finally prove the optimum over the full class. The stationary
effective upper-link coefficient satisfies
\(0\leq k_{\rm eff}[P]\leq A_m\), with equality at the upper bound precisely
when \(\langle e^{i\alpha}\rangle_P=0\). The same condition maximizes the
stationary cooling power in the refrigeration window. Coherence is essential
to the microscopic penalty, although the stationary population balance after
coherence elimination has an exact classical-network representation. The
result is therefore a constrained control theorem, not a claim of a uniquely
quantum stationary advantage. Phase cycling changes the cycle current but not
the energy quanta per cycle; a device-level coefficient of performance would
require a microscopic energetic model of the measuring apparatus.

\section{General control framework and minimal phase-cycling realization}
\label{sec:model}

\begin{figure}[t]
\centering
\begin{overpic}[width=0.95\columnwidth]{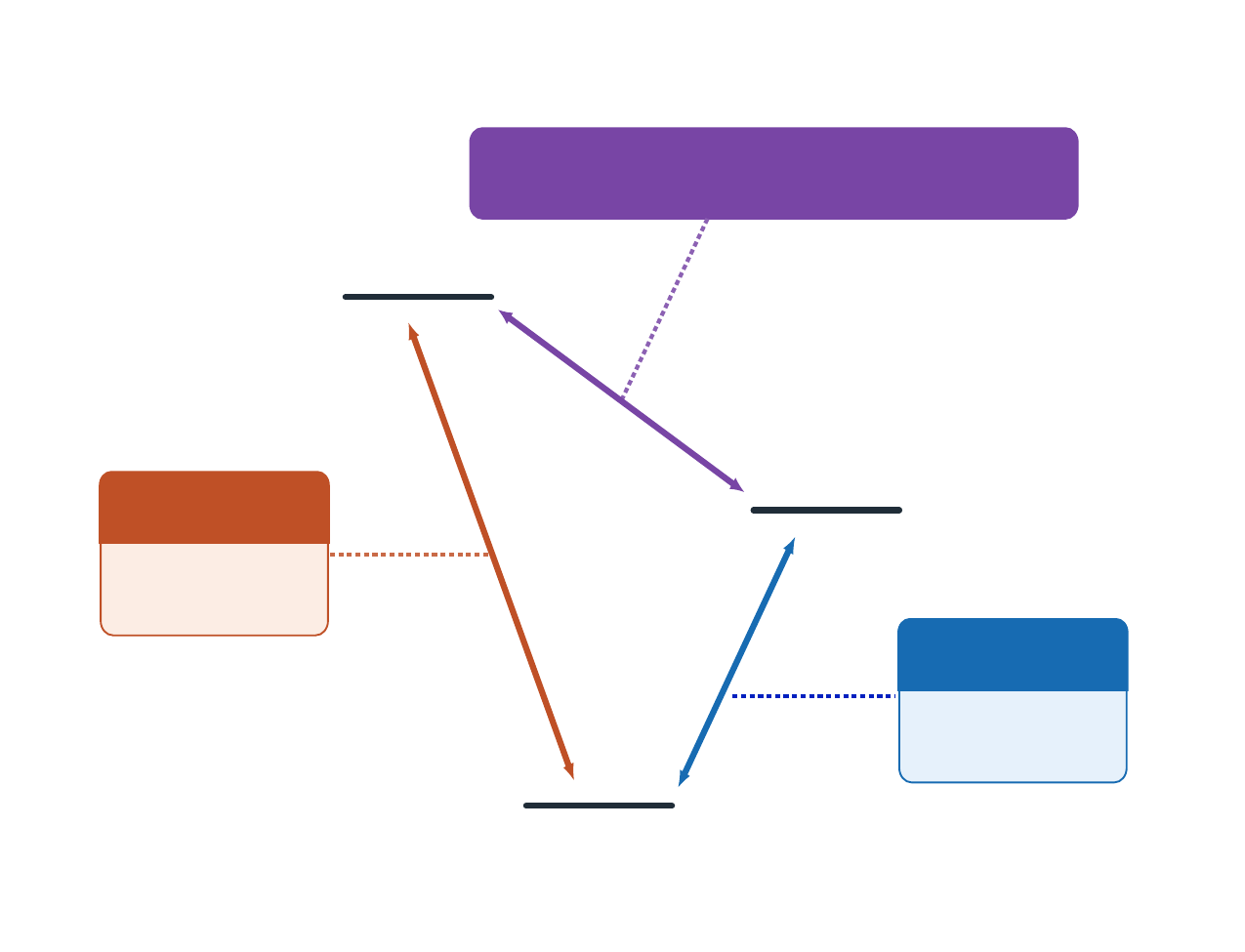}

\put(62,63){\makebox(0,0){%
  \color{white}\scriptsize\bfseries Nonselective measurements}}

\put(61,52){\makebox(0,0){%
  \color{modelpurple}\footnotesize $\dot E_{\mathrm{meas}}$}}

\put(24,53){\makebox(0,0){%
  \color{modelblue}\footnotesize $|3\rangle$}}
\put(75,36){\makebox(0,0){%
  \color{modelblue}\footnotesize $|2\rangle$}}
\put(39,12){\makebox(0,0){%
  \color{modelblue}\footnotesize $|1\rangle$}}

\put(32,55){\makebox(0,0){%
  \color{modelblue}\footnotesize $\Delta_{31}$}}
\put(66,38){\makebox(0,0){%
  \color{modelblue}\footnotesize $\Delta_{21}$}}
\put(47,9){\makebox(0,0){%
  \color{modelblue}\footnotesize $0$}}

\put(16,36){\makebox(0,0){%
  \color{white}\scriptsize\bfseries Hot bath}}
\put(17,30){\makebox(0,0){%
  \color{modelhot}\footnotesize $T_h$}}

\put(34,35){\makebox(0,0){%
  \color{modelhot}\footnotesize $\dot Q_h$}}

\put(81,24){\makebox(0,0){%
  \color{white}\scriptsize\bfseries Cold bath}}
\put(82,18){\makebox(0,0){%
  \color{modelcold}\footnotesize $T_c$}}

\put(64,24){\makebox(0,0){%
  \color{modelcold}\footnotesize $\dot Q_c$}}

\end{overpic}
\caption{Schematic of the measurement-assisted three-level refrigerator.
The cold bath at temperature \(T_c\) couples
\(\ket{1}\leftrightarrow\ket{2}\), the hot bath at \(T_h\) couples
\(\ket{1}\leftrightarrow\ket{3}\), and nonselective measurements act on
\(\ket{2}\leftrightarrow\ket{3}\).
Solid double-headed arrows represent the three transition channels;
dashed lines connect each channel to the corresponding bath or
measurement operation.
The energy currents \(\dot Q_c\), \(\dot Q_h\), and
\(\dot E_{\rm meas}\) are defined as positive into the working medium.
At each measurement event, one of two azimuthal phases is independently
selected with equal probability. No measurement outcome is recorded
or used for feedback.}
\label{fig:model}
\end{figure}

\subsection{Three-level working medium}

We consider the measurement-assisted three-level refrigerator shown in
Fig.~\ref{fig:model}. The working medium has energy eigenstates
\(\{\ket{1},\ket{2},\ket{3}\}\). We choose \(E_1=0\) and write
\(E_2=\Delta_{21}\) and \(E_3=\Delta_{31}\), with
\(0<\Delta_{21}<\Delta_{31}\). The energy gap between the two upper
levels is \(\Delta_{32}\equiv\Delta_{31}-\Delta_{21}\).
In units \(\hbar=k_B=1\), the Hamiltonian is
\begin{equation}
H_S=\Delta_{21}\ket{2}\bra{2}+\Delta_{31}\ket{3}\bra{3}.
\label{eq:HS}
\end{equation}
The cold reservoir at temperature \(T_c\) couples
\(\ket{1}\leftrightarrow\ket{2}\), whereas the hot reservoir at
\(T_h\) couples \(\ket{1}\leftrightarrow\ket{3}\).
Nonselective measurements act on the upper transition
\(\ket{2}\leftrightarrow\ket{3}\).
We denote the heat currents from the cold and hot reservoirs by
\(\dot Q_c\) and \(\dot Q_h\), respectively, and the energy current
from the measurement channel by \(\dot E_{\rm meas}\).
All currents are defined as positive into the working medium.

For independent, spectrally selective reservoirs in the
weak-coupling Markovian and secular regime, the local bath
generators take the standard Lindblad
form~\cite{Gorini1976,Lindblad1976,Breuer2002}:
\begin{equation}
\mathcal L_c[\rho]=\lambda_c\mathcal D[\ket{1}\bra{2}]\rho
+\bar\lambda_c\mathcal D[\ket{2}\bra{1}]\rho,
\label{eq:Lc}
\end{equation}
\begin{equation}
\mathcal L_h[\rho]=\lambda_h\mathcal D[\ket{1}\bra{3}]\rho
+\bar\lambda_h\mathcal D[\ket{3}\bra{1}]\rho,
\label{eq:Lh}
\end{equation}
where \(\mathcal D[A]\rho\equiv A\rho A^\dagger-
\{A^\dagger A,\rho\}/2\). The downward rates are \(\lambda_c\) and
\(\lambda_h\), and local detailed balance fixes the excitation rates as
\(\bar\lambda_c=\lambda_c e^{-\Delta_{21}/T_c}\) and
\(\bar\lambda_h=\lambda_h e^{-\Delta_{31}/T_h}\). The representative
parameters used below keep the bath rates well below the relevant Bohr
frequencies at the plotted and optimized operating points.

The two reservoirs share the state \(\ket{1}\), but their jump channels are
independent. Taking the \((2,3)\) matrix element of
\(-i[H_S,\rho]+\mathcal L_c[\rho]+\mathcal L_h[\rho]\), with
Eqs.~\eqref{eq:HS}--\eqref{eq:Lh}, gives
\begin{equation}
\left.\dot\rho_{23}\right|_{H_S+c+h}
=\left(i\Delta_{32}-\frac{\lambda_c+\lambda_h}{2}\right)\rho_{23}.
\label{eq:bath-coherence}
\end{equation}
Equation~\eqref{eq:bath-coherence} is derived explicitly in
Appendix~\ref{app:bath}. The Hamiltonian rotates the upper-state coherence at frequency
\(\Delta_{32}\), while the two downward channels supply the dephasing rate
\((\lambda_c+\lambda_h)/2\). The excitation terms do not contribute to this
matrix element. Equation~\eqref{eq:bath-coherence} is homogeneous: the
independent reservoirs can rotate and damp a preexisting \(\rho_{23}\), but
they cannot generate a source proportional to the population imbalance.

\subsection{One tilted measurement: two coupled actions}
\label{sec:single-measurement}

The nonselective measurement acts in
\(\mathcal H_{23}=\operatorname{span}\{\ket{2},\ket{3}\}\). Its polar
angle \(\theta\) fixes the tilt of the measured axis relative to the energy
basis, and its azimuthal phase \(\alpha\) fixes the direction of that axis in
the upper-state coherence plane. We use the orthonormal measurement states
\begin{subequations}\label{eq:measurement-basis}
\begin{align}
\ket{\chi_+(\alpha)}
&=\cos\frac{\theta}{2}\ket{2}
+e^{i\alpha}\sin\frac{\theta}{2}\ket{3},
\label{eq:chi-plus}\\
\ket{\chi_-(\alpha)}
&=-e^{-i\alpha}\sin\frac{\theta}{2}\ket{2}
+\cos\frac{\theta}{2}\ket{3}.
\label{eq:chi-minus}
\end{align}
\end{subequations}
With \(\Pi_\pm(\alpha)=\ket{\chi_\pm(\alpha)}\bra{\chi_\pm(\alpha)}\)
and \(\Pi_1=\ket{1}\bra{1}\), one outcome-discarded event applies the
channel
\begin{equation}
\mathcal M_{\theta,\alpha}[\rho]
=\Pi_1\rho\Pi_1+\sum_{s=\pm}\Pi_s(\alpha)\rho\Pi_s(\alpha).
\label{eq:Mtheta}
\end{equation}
Here an event means one application of Eq.~\eqref{eq:Mtheta}. The sum over
\(s\) is the quantum average over the unrecorded outcomes of that event; it
does not yet average over the event time or over different choices of
\(\alpha\).

The azimuthal dependence can be exposed by the relative phase rotation
\(V_\alpha\equiv\exp\{i\alpha(\ket{3}\bra{3}-\ket{2}\bra{2})/2\}\).
It multiplies \(\ket{2}\) and \(\ket{3}\) by opposite phases without
changing their populations. Applying it to the zero-phase basis in
Eqs.~\eqref{eq:chi-plus} and~\eqref{eq:chi-minus} gives the basis at
\(\alpha\), up to physically irrelevant overall phases. Consequently,
\begin{equation}
\mathcal M_{\theta,\alpha}
=\mathcal V_\alpha\circ\mathcal M_{\theta,0}\circ\mathcal V_{-\alpha},
\label{eq:phase-covariance}
\end{equation}
where \(\mathcal V_\alpha[\rho]\equiv V_\alpha\rho V_\alpha^\dagger\).
Because \([V_\alpha,H_S]=0\), the phase rotations alone leave the level
populations and the system energy unchanged. Equation~\eqref{eq:phase-covariance}
therefore shows that changing \(\alpha\) rotates the measurement direction
without changing \(\theta\) or the energy-basis transition strength. It also
provides an implementation picture: the same fixed tilted measurement can be
combined with programmable relative-phase rotations, without any
outcome-conditioned operation. This distinction is relevant when the
system--meter interaction fixes the polar angle, while a relative \(z\)-phase
can be changed by a control pulse or frame update. The energetic and temporal
cost of that phase operation is hardware dependent and is not included in the
reduced map below.

Measurement events are treated as effectively instantaneous and form a
Poisson process with total rate \(\gamma_m\). During a short interval
\(dt\), one event occurs with probability \(\gamma_mdt+O(dt^2)\), whereas
two or more events have probability \(O(dt^2)\). Averaging over these random
event times for one fixed phase gives
\begin{equation}
\mathcal L_{m,\alpha}[\rho]
=\gamma_m\bigl(\mathcal M_{\theta,\alpha}[\rho]-\rho\bigr).
\label{eq:fixed-phase-generator}
\end{equation}
The corresponding mean waiting time is \(1/\gamma_m\).
This map-valued jump construction is the Poissonian counterpart of
repeated-interaction descriptions of open-system dynamics
~\cite{Strasberg2017,Ciccarello2022}.

We now separate the two physical actions of one measurement event. The
nonselective map in Eq.~\eqref{eq:Mtheta} already includes the sum over
the discarded outcomes. Starting from the energy eigenstate
\(\ket{2}\bra{2}\), the population transferred to \(\ket{3}\) by one
event is
\( 
P_{2\to3}
\equiv
\bra{3}
\mathcal M_{\theta,\alpha}
\!\left[\ket{2}\bra{2}\right]
\ket{3}
=
\sum_{s=\pm}
\left|\braket{\chi_s(\alpha)|2}\right|^2
\left|\braket{3|\chi_s(\alpha)}\right|^2
=
\frac12\sin^2\theta .
\) 
Interchanging the two upper states gives the same probability in the
opposite direction, \(P_{3\to2}=P_{2\to3}\). Since the measurement
events occur according to a Poisson process with rate \(\gamma_m\), the
corresponding continuous-time direct mixing rate is
\begin{equation}
A_m
\equiv
\gamma_m P_{2\to3}
=
\frac{\gamma_m}{2}\sin^2\theta .
\label{eq:direct-mixing}
\end{equation}
Here, \(P_{2\to3}\) is the dimensionless population-transfer probability
per measurement event, whereas \(A_m\) is the transition rate entering
the ensemble-averaged population equations. Neither quantity depends on
the azimuthal phase \(\alpha\), because the direct population transfer
contains only squared projection amplitudes.

In addition to transferring population between the two upper states, the
same nonselective measurement event can generate upper-state coherence.
To isolate the coherence source produced by the population imbalance, we
consider a state for which the upper-state coherence vanishes immediately
before the event, \(\rho_{23}=0\), while the populations \(p_2\) and
\(p_3\) remain arbitrary. Substituting the projectors from
Eqs.~\eqref{eq:chi-plus} and~\eqref{eq:chi-minus} into the nonselective
map in Eq.~\eqref{eq:Mtheta} gives
\begin{equation}
\bra{2}\mathcal M_{\theta,\alpha}[\rho]\ket{3}
=
\frac12\sin\theta\cos\theta\,
e^{-i\alpha}(p_2-p_3).
\label{2m3}
\end{equation}
This relation describes the coherence generated by one nonselective
measurement event; it is not a continuous-time equation. Since such
events occur according to a Poisson process with rate \(\gamma_m\),
multiplying the single-event change by \(\gamma_m\) gives the
population-dependent coherence source in the ensemble-averaged equation
for \(\rho_{23}\).

The source magnitude is determined by the polar angle \(\theta\) and the
population imbalance \(p_2-p_3\), whereas its orientation in the
\((\operatorname{Re}\rho_{23},\operatorname{Im}\rho_{23})\) plane is
controlled by the azimuthal phase \(\alpha\). Thus, for
\(0<\theta<\pi/2\), one tilted nonselective measurement channel combines
a phase-independent population-transfer process with a phase-tagged
coherence source. The projector multiplication is worked out in
Appendix~\ref{app:fixed-map}, where Eq.~\eqref{2m3} is extended to a state
containing preexisting upper-state coherence.

\subsection{General random-phase control class}
\label{sec:protocol}

We now promote \(\alpha\) from a fixed parameter to an externally prescribed
random control variable. Immediately before the \(n\)th measurement event, a
phase \(\alpha_n\) is selected independently from a normalized distribution
\(P(\alpha)\), with \(\int_0^{2\pi}P(\alpha)d\alpha=1\). The choice is
independent of the system state, all earlier phase choices, and all
measurement outcomes. The protocol is therefore open loop and contains no
measurement-based feedback.

For a prescribed realization, an event at \(t_n\) updates the conditional
state as
\(\rho(t_n^+)=\mathcal M_{\theta,\alpha_n}[\rho(t_n^-)]\). Averaging over the
Poisson event times and the independent phase choices gives
\begin{equation}
\mathcal L_P[\rho]
=
\gamma_m
\left[
\int_0^{2\pi}d\alpha\,P(\alpha)
\mathcal M_{\theta,\alpha}[\rho]
-\rho
\right].
\label{eq:LP}
\end{equation}
A detailed derivation is given in
Appendix~\ref{app:random-phase-generator}. The map
\(\mathcal M_{\theta,\alpha}\) already averages over the discarded quantum
outcomes; Eq.~\eqref{eq:LP} performs the additional classical average over
event times and phases. Henceforth \(\rho\) denotes the unconditional state.

Combining Eq.~\eqref{eq:LP} with Eqs.~\eqref{eq:HS}--\eqref{eq:Lh} gives the
full reduced dynamics
\begin{equation}
\dot\rho=-i[H_S,\rho]+\mathcal L_c[\rho]
+\mathcal L_h[\rho]+\mathcal L_P[\rho].
\label{eq:master}
\end{equation}
The master equation is time homogeneous because the event rate and the phase
distribution are fixed and independent of the state. Equation~\eqref{eq:master}
defines the complete reduced-map control class studied in this work: the bath
parameters, level spacings, total event rate \(\gamma_m\), polar angle
\(\theta\), and outcome-discarding rule are held fixed, while only
\(P(\alpha)\) is varied. Temporally correlated or deterministic phase
sequences are outside this independent-sampling control class.

If \(k_{\rm eff}[P]\) denotes the
stationary coefficient relating the measurement-induced upper-link current to
the stationary population imbalance, then
\[
0\leq k_{\rm eff}[P]\leq A_m,
\qquad
k_{\rm eff}[P]=A_m
\ \Longleftrightarrow\ 
\left\langle e^{i\alpha}\right\rangle_P=0
\]
for \(\gamma_m>0\) and \(0<\theta<\pi/2\). The first phase moment is the
resultant of the phase-tagged coherence-source vectors. Its vanishing is
therefore the natural general condition for removing the coherent feedback
without altering the direct mixing. Before proving this statement for an
arbitrary distribution, we analyze the smallest family that makes its physical
content transparent.

\subsection{Minimal balanced two-phase realization}
\label{sec:minimal-two-phase}

The simplest nontrivial phase-cycling protocol uses two phase settings,
\(\alpha_n=+\delta/2\) and \(\alpha_n=-\delta/2\), selected independently
with equal probability:
\( 
\Pr\!\left(\alpha_n=+\delta/2\right)
=
\Pr\!\left(\alpha_n=-\delta/2\right)
=
\frac{1}{2}.
\) 
Equivalently, the discrete phase ensemble is represented by
\( 
P_{\rm pc}(\alpha)
=
\frac{1}{2}
\left[
\delta_{2\pi}(\alpha-\delta/2)
+
\delta_{2\pi}(\alpha+\delta/2)
\right],
\)
where \(\delta_{2\pi}\) denotes the \(2\pi\)-periodic Dirac delta function.
Substitution into Eq.~\eqref{eq:LP} reduces the phase integral to the
average over the two settings and gives
\begin{equation}
\mathcal L_{\rm pc}[\rho]
=
\frac{\gamma_m}{2}
\left[
\mathcal M_{\theta,+\delta/2}[\rho]
+
\mathcal M_{\theta,-\delta/2}[\rho]
-
2\rho
\right].
\label{eq:Lpc}
\end{equation}
It replaces
\(\mathcal L_P\) for this particular phase distribution and is not an
additional dynamical term. 

This balanced two-phase ensemble is a minimal one-parameter slice through
Eq.~\eqref{eq:LP}, not an assumed global optimizer. The event rate, polar
angle, and single-event population transfer remain unchanged as \(\delta\) is
varied, whereas the coherence source is multiplied by
\(
\tfrac12(e^{-i\delta/2}+e^{i\delta/2})=\cos(\delta/2)
\).
Thus \(\delta=0\) reproduces one fixed phase, whereas \(\delta=\pi\) cancels
the two opposite source vectors without changing the population-transfer
channel.

\begin{figure*}[!t]
\centering
\hspace*{0.025\textwidth}%
\begin{minipage}{0.43\textwidth}
\centering
\begin{overpic}[width=\linewidth]{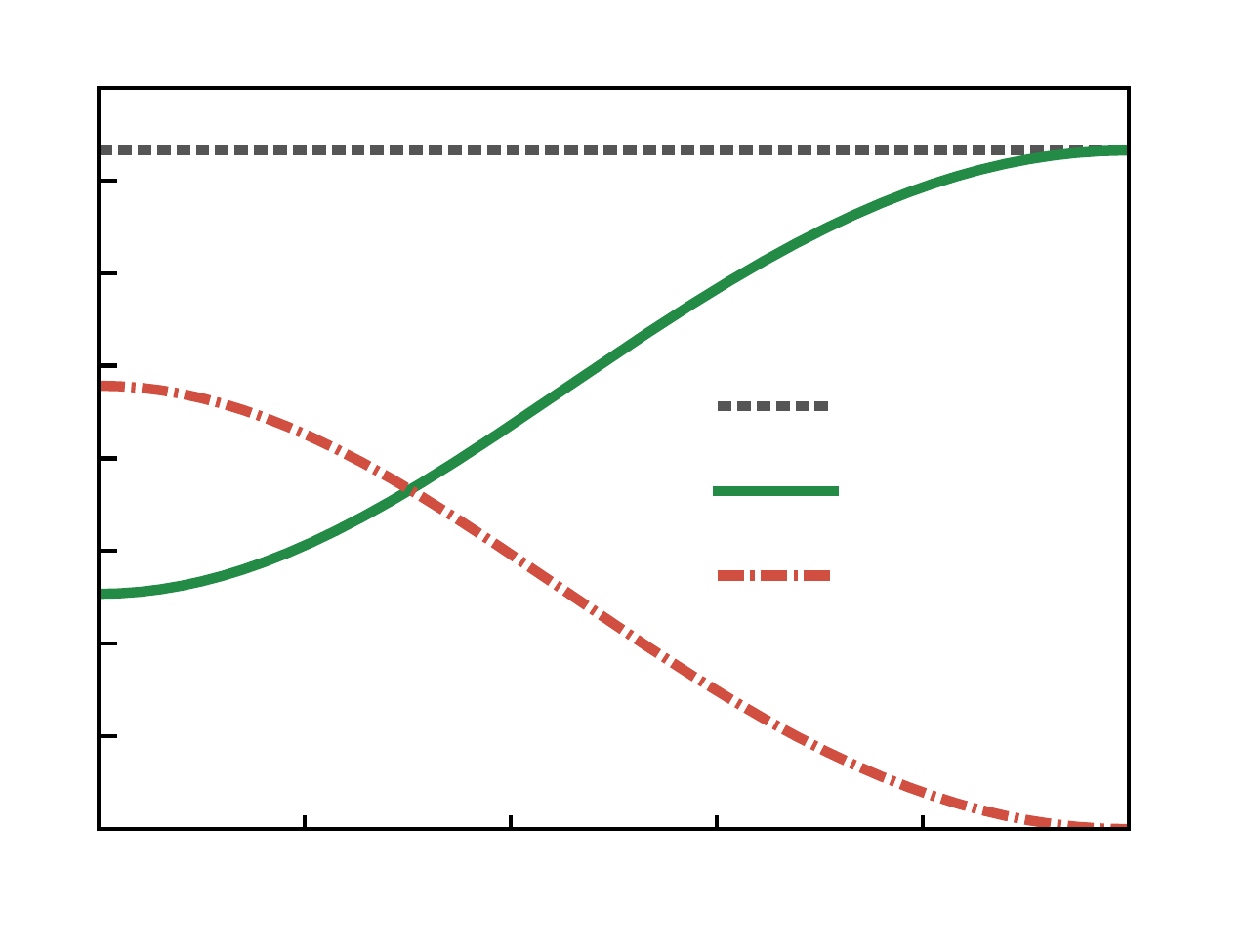}

\put(8,71){{(a)}}

\put(32,2){\scalebox{1.15}{phase separation $\delta/\pi$}}
\put(-6,36){\rotatebox{90}{\scalebox{1.15}{rate}}}

\put(7,6){\scalebox{1.0}{$0.0$}}
\put(22,6){\scalebox{1.0}{$0.2$}}
\put(39,6){\scalebox{1.0}{$0.4$}}
\put(56,6){\scalebox{1.0}{$0.6$}}
\put(73,6){\scalebox{1.0}{$0.8$}}
\put(88,6){\scalebox{1.0}{$1.0$}}

\put(-1,9){\scalebox{1.0}{$0.00$}}
\put(-1,24){\scalebox{1.0}{$0.02$}}
\put(-1,39){\scalebox{1.0}{$0.04$}}
\put(-1,53){\scalebox{1.0}{$0.06$}}
\put(-1,68){\scalebox{1.0}{$0.08$}}

\put(68,43){\scalebox{1.0}{$A_m$}}
\put(68,36){\scalebox{1.0}{$k_{\rm eff}$}}
\put(68,29){\scalebox{1.0}{$\Delta k_{\rm coh}$}}

\end{overpic}
\end{minipage}%
\hfill
\begin{minipage}{0.43\textwidth}
\centering
\begin{overpic}[width=\linewidth]{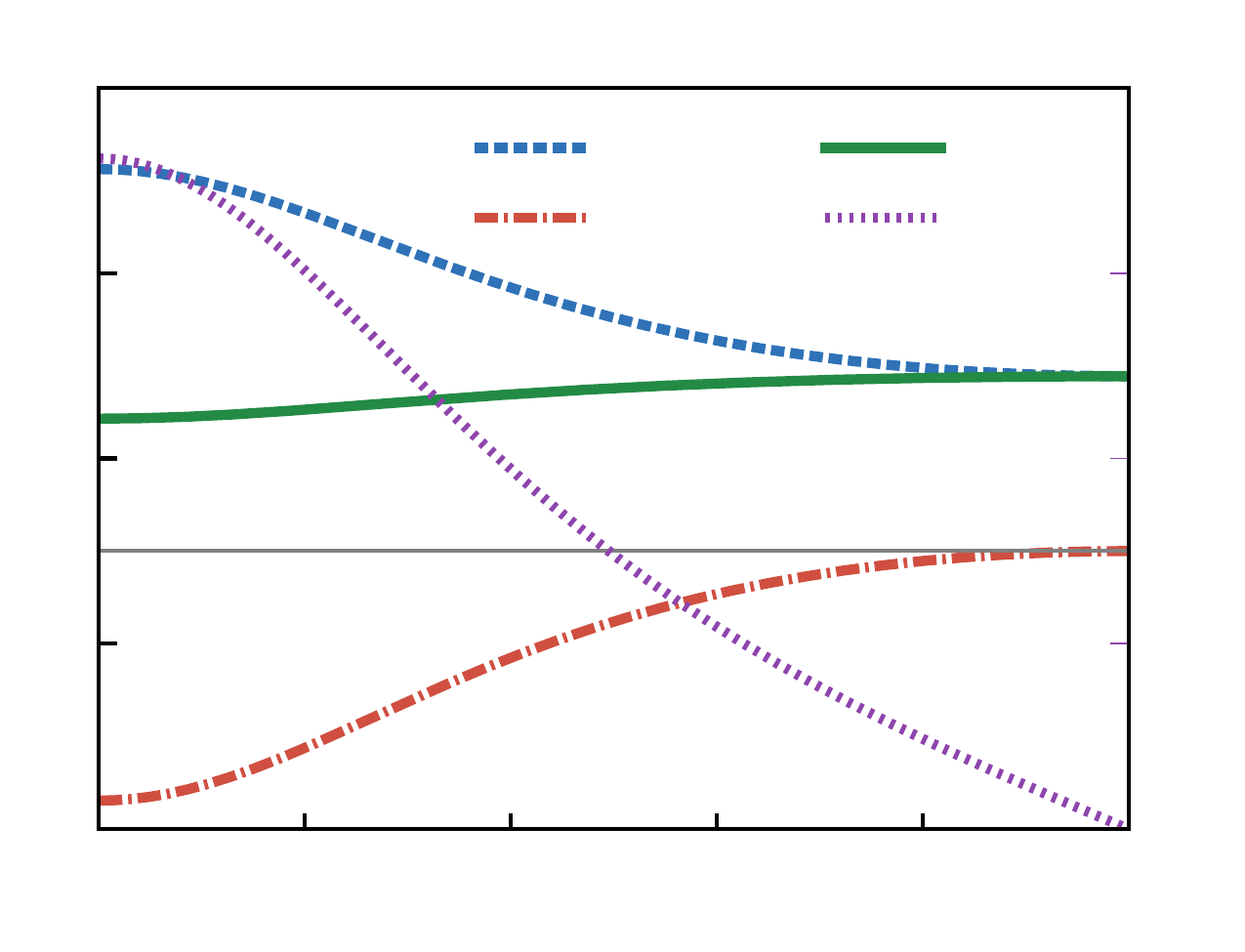}

\put(8,71){{(b)}}

\put(35,2){\scalebox{1.15}{phase separation $\delta/\pi$}}
\put(-10,20){\rotatebox{90}{\scalebox{1.15}{stationary current}}}
\put(102,35){\rotatebox{90}{\scalebox{1.15}{$C_{\ell_1}^{\rm ss}$}}}

\put(7,6){\scalebox{1.0}{$0.0$}}
\put(22,6){\scalebox{1.0}{$0.2$}}
\put(39,6){\scalebox{1.0}{$0.4$}}
\put(56,6){\scalebox{1.0}{$0.6$}}
\put(73,6){\scalebox{1.0}{$0.8$}}
\put(88,6){\scalebox{1.0}{$1.0$}}

\put(-6,9){\scalebox{1.0}{$-0.003$}}
\put(-6,23){\scalebox{1.0}{$-0.001$}}
\put(-4,38){\scalebox{1.0}{$0.001$}}
\put(-4,53){\scalebox{1.0}{$0.003$}}
\put(-4,68){\scalebox{1.0}{$0.005$}}

\put(91,10){\scalebox{1.0}{$0.000$}}
\put(91,24){\scalebox{1.0}{$0.005$}}
\put(91,39){\scalebox{1.0}{$0.010$}}
\put(91,53){\scalebox{1.0}{$0.015$}}
\put(91,68){\scalebox{1.0}{$0.020$}}

\put(47,64){\scalebox{1.0}{$J_{\rm dir}$}}
\put(47,58){\scalebox{1.0}{$J_{\rm fb}$}}
\put(76,64){\scalebox{1.0}{$J_m$}}
\put(76,58){\scalebox{1.0}{$C_{\ell_1}^{\rm ss}$}}

\end{overpic}
\end{minipage}%
\hspace*{0.025\textwidth}%
\par
\caption{Removal of coherence-mediated feedback at
\(\Delta_{21}=0.34\). (a) Bare direct mixing rate \(A_m\), stationary effective coefficient
\(k_{\rm eff}\), and feedback penalty
\(\Delta k_{\rm coh}=A_m-k_{\rm eff}\). Phase cycling leaves \(A_m\) fixed
and removes only the penalty. (b) Direct term
\(J_{\rm dir}=A_m\Delta p^{\rm ss}\), feedback correction
\(J_{\rm fb}=-B_m c_\delta R^{\rm ss}\), total measurement current
\(J_m=J_{\rm dir}+J_{\rm fb}\), and stationary coherence
\(C_{\ell_1}^{\rm ss}=2|\rho_{23}^{\rm ss}|\). Parameters are
\(T_c=0.6\), \(T_h=0.8\), \(\Delta_{31}=1\),
\(\lambda_c=\lambda_h=0.03\), \(\gamma_m=1\), and
\(\theta=\pi/8\). Energies and rates are expressed in units of
\(\Delta_{31}\).}
\label{fig:mechanism}
\end{figure*}
\section{Minimal two-phase mechanism: cancellation of coherence-mediated feedback}
\label{sec:mechanism}

\subsection{Closed population--coherence dynamics}

We now specialize the general control class to the balanced two-phase family
introduced in Sec.~\ref{sec:minimal-two-phase}. This specialization is used to
expose the microscopic pathway by which phase cycling restores the direct
population mixing. Write
\(R\equiv\operatorname{Re}\rho_{23}\),
\(I\equiv\operatorname{Im}\rho_{23}\), and
\(\Delta p\equiv p_2-p_3\). We also use
\(B_m\equiv\gamma_m\sin\theta\cos\theta\),
\(C_m\equiv B_m/2\),
\(c_\delta\equiv\cos(\delta/2)\), and
\(s_\delta\equiv\sin(\delta/2)\). When 
\(\alpha_\pm=\pm\delta/2\), we can obtain closed equations: 
\begin{subequations}\label{eq:population-coherence}
\begin{align}
\dot p_2&=\bar\lambda_c p_1-\lambda_c p_2
-A_m\Delta p+B_m c_\delta R,
\label{eq:pdot2}\\
\dot p_3&=\bar\lambda_h p_1-\lambda_h p_3
+A_m\Delta p-B_m c_\delta R,
\label{eq:pdot3}\\
\dot R&=-\Delta_{32}I-\Gamma_R(\delta)R
+C_m c_\delta\Delta p,
\label{eq:Rdot}\\
\dot I&=\Delta_{32}R-\Gamma_I(\delta)I.
\label{eq:Idot}
\end{align}
\end{subequations}
Here \(\Gamma_R(\delta)\equiv\gamma_\varphi+
\gamma_m[1-\sin^2\theta\,c_\delta^2]\),
\(\Gamma_I(\delta)\equiv\gamma_\varphi+
\gamma_m[1-\sin^2\theta\,s_\delta^2]\), and
\(\gamma_\varphi\equiv(\lambda_c+\lambda_h)/2\). The coupled equations are
obtained from the two-phase average in Appendix~\ref{app:two-phase}.

Equations~\eqref{eq:population-coherence} contain the microscopic mechanism
of the paper. The terms \(\mp A_m\Delta p\) in
Eqs.~\eqref{eq:pdot2} and~\eqref{eq:pdot3} are the useful direct population
mixing that closes the refrigeration cycle. The same imbalance enters
Eq.~\eqref{eq:Rdot} through \(C_m c_\delta\Delta p\), thereby generating a
real upper-state coherence. That coherence returns to the population
equations through \(\pm B_m c_\delta R\). The two terms therefore form a
closed population-to-coherence-to-population feedback path. During this
process, the Hamiltonian terms proportional to \(\Delta_{32}\) rotate the
coherence between \(R\) and \(I\), while \(\Gamma_R\) and \(\Gamma_I\)
represent the combined reservoir- and measurement-induced damping of the two
coherence quadratures.

The phase separation acts selectively on this feedback loop. The direct rate
\(A_m\) does not depend on \(\delta\), because the single-event population
transfer is phase independent. In contrast, both the source
\(C_m c_\delta\Delta p\) and the return term \(B_m c_\delta R\) carry the
factor \(c_\delta\). Phase cycling therefore does not replace the original
measurement by a stronger channel; it progressively removes the coherent
path generated by the same channel.

The stationary coherence can be eliminated directly from
Eqs.~\eqref{eq:Rdot} and~\eqref{eq:Idot}. Setting \(\dot I=0\) gives
\(I^{\rm ss}=\Delta_{32}R^{\rm ss}/\Gamma_I(\delta)\), and substitution into
the stationary \(R\) equation gives
\(R^{\rm ss}=C_m c_\delta\Delta p^{\rm ss}/D_\delta\), with
\(D_\delta\equiv\Gamma_R(\delta)+\Delta_{32}^2/\Gamma_I(\delta)>0\).
Inserting this stationary response into the measurement part of
Eq.~\eqref{eq:pdot2} yields
\(J_m=k_{\rm eff}(\delta)\Delta p^{\rm ss}\), where
\(J_m\equiv A_m\Delta p^{\rm ss}-B_m c_\delta R^{\rm ss}\). Thus
\(k_{\rm eff}(\delta)\) is the stationary effective upper-link coefficient
relating the measurement-induced current to \(\Delta p^{\rm ss}\); it is not
an assertion that the coherence-eliminated population dynamics is Markovian at
all times. The coefficient is
\begin{equation}
k_{\rm eff}(\delta)=A_m-\Delta k_{\rm coh}(\delta).
\label{eq:keff}
\end{equation}
Here \(\Delta k_{\rm coh}(\delta)\equiv B_mC_m c_\delta^2/D_\delta\geq0\)
is the coherence-feedback penalty. The first term in
Eq.~\eqref{eq:keff} is the bare direct mixing rate, whereas the subtracted
term is the part of that response canceled after the same measurement
channel generates, rotates, damps, and feeds back the upper-state coherence.
It is therefore a correction to the same measurement-induced current, not an
independent current and not an additional thermodynamic resource. The
elimination steps are provided in Appendix~\ref{app:two-phase}, together
with a proof that \(\Delta k_{\rm coh}\) decreases monotonically as
\(\delta\) increases from zero to \(\pi\).

At the balanced \(\pi\)-separated point, \(c_\delta=0\), so
\(R^{\rm ss}=I^{\rm ss}=0\) and \(k_{\rm eff}(\pi)=A_m\). This result is
stronger than a cancellation that appears only after a stationary state has
been solved. Setting \(c_\delta=0\) directly in
Eqs.~\eqref{eq:population-coherence} removes both directions of the
population--coherence coupling. The full generator is then block diagonal
between the population sector and the upper-state-coherence sector. An
initially present \(\rho_{23}\) may still precess and decay, but it cannot
modify the populations or the cooling current.

\subsection{Rate-level and current-level picture}

For the numerical discussion, we decompose the stationary measurement-induced
upper-state current as \(J_m=J_{\rm dir}+J_{\rm fb}\), where
\(J_{\rm dir}\equiv A_m\Delta p^{\rm ss}\) is the direct population-mixing
term and \(J_{\rm fb}\equiv-B_m c_\delta R^{\rm ss}\) is the
coherence-mediated feedback correction. They are algebraic contributions to
the same current, not independent thermodynamic currents. The only
independent nonvanishing stationary off-diagonal element is
\(\rho_{23}^{\rm ss}\), with
\(\rho_{32}^{\rm ss}=(\rho_{23}^{\rm ss})^*\). We quantify its magnitude by
the steady-state \(\ell_1\)-norm coherence
\(C_{\ell_1}^{\rm ss}\equiv\sum_{i\neq j}|\rho_{ij}^{\rm ss}|=
2|\rho_{23}^{\rm ss}|\)~\cite{Baumgratz2014}; more generally, coherence in
thermodynamic processes carries constraints that are not captured by
population free-energy data alone~\cite{Lostaglio2015}.

Figure~\ref{fig:mechanism}(a) displays the mechanism at the rate level. The
horizontal \(A_m\) curve is the fixed-resource benchmark. At \(\delta=0\),
coherence feedback gives the largest penalty. Increasing \(\delta\) turns the
two source vectors toward opposite directions, and both links of the feedback
path decrease with \(c_\delta\). At \(\delta=\pi\),
\(\Delta k_{\rm coh}=0\) and \(k_{\rm eff}=A_m\). The rise of
\(k_{\rm eff}\) is therefore a recovery of the existing direct rate, not an
increase of measurement strength.

The current decomposition in Fig.~\ref{fig:mechanism}(b) shows the same
mechanism after the stationary populations have adjusted self-consistently. For \(\delta=0\), the algebraic
direct term is opposed by \(J_{\rm fb}<0\). As \(\delta\) increases,
\(C_{\ell_1}^{\rm ss}\) and the negative correction vanish. The direct term
also decreases because the restored upper-state rate relaxes
\(\Delta p^{\rm ss}\), but the removal of \(J_{\rm fb}\) is larger, so
\(J_m\) increases. At \(\delta=\pi\), \(J_{\rm fb}=0\) and
\(J_m=J_{\rm dir}\). The relevant return variable is the phase-aligned
component \(c_\delta R^{\rm ss}\); the plotted coherence shows its
simultaneous removal in the symmetric protocol.

\section{General theorem and cooling-power bound}
\label{sec:optimality}

\subsection{Arbitrary phase distributions: exact rate bound}
\label{sec:global-optimality}

We now optimize over every normalized independent phase distribution
\(P(\alpha)\) in Eq.~\eqref{eq:LP}, including unequal, multiphase, and
continuous distributions, while keeping \(\gamma_m\), \(\theta\), the bath
parameters, and the level spacings fixed.

Averaging the fixed-phase equations introduces the first phase moment
\(\bm m\equiv
(\langle\cos\alpha\rangle_P,-\langle\sin\alpha\rangle_P)^T\), where
\(\langle f\rangle_P\equiv
\int_0^{2\pi}d\alpha\,P(\alpha)f(\alpha)\). For a given phase, the
population imbalance generates a coherence-source vector proportional to
\((\cos\alpha,-\sin\alpha)^T\Delta p\). The vector \(\bm m\) is therefore
the resultant of these phase-dependent sources after averaging. A nonzero
\(\bm m\) represents a residual population-to-coherence source, whereas
\(\bm m=0\) means that the sources cancel at the generator level.

The same phase average produces a symmetric positive-definite damping matrix
\(\bm\Gamma_P\), determined by the second phase moments. Hamiltonian
precession in the upper-state coherence plane is represented by
\(\bm\Omega\equiv
\Delta_{32}\bigl(\begin{smallmatrix}0&-1\\1&0\end{smallmatrix}\bigr)\).
Writing \(\bm u=(R,I)^T\), the averaged coherence equation takes the form
\(\dot{\bm u}=(-\bm\Gamma_P+\bm\Omega)\bm u+
C_m\bm m\Delta p\), while the measurement contribution to the
equation for \(p_2\) is
\(-A_m\Delta p+B_m\bm m^T\bm u\).

At stationarity, the coherence vector is
\begin{equation}
\bm u^{\rm ss}
=
C_m(\bm\Gamma_P-\bm\Omega)^{-1}
\bm m\,\Delta p^{\rm ss}.
\label{eq:general-stationary-coherence}
\end{equation}
Defining the stationary effective upper-link coefficient through
\(J_m=k_{\rm eff}[P]\Delta p^{\rm ss}\), substitution of
Eq.~\eqref{eq:general-stationary-coherence} into the population current gives
\begin{equation}
k_{\rm eff}[P]
=
A_m
-
B_mC_m\,\bm m^T
(\bm\Gamma_P-\bm\Omega)^{-1}\bm m.
\label{eq:general-keff}
\end{equation}
The phase-averaged equations and
Eqs.~\eqref{eq:general-stationary-coherence} and~\eqref{eq:general-keff}
are derived in Appendix~\ref{app:general-phase}, where the explicit form
of \(\bm\Gamma_P\) is also specified.

In Eq.~\eqref{eq:general-keff}, \(\bm m\) is the residual coherence source,
the inverse matrix is its stationary response, and the final contraction is
the return to the populations. The subtracted quadratic form is therefore the
feedback penalty for an arbitrary phase distribution.

This quadratic correction is non-negative and cannot exceed \(A_m\).
Consequently, every normalized open-loop phase distribution obeys
\begin{equation}
0\leq k_{\rm eff}[P]\leq A_m.
\label{eq:rate-bound}
\end{equation}
The upper limit is the ceiling within the present control class at fixed
\(\gamma_m\) and \(\theta\): phase statistics cannot create more direct
mixing than the single-event map already contains.

For \(\gamma_m>0\) and the nontrivial tilted regime
\(0<\theta<\pi/2\), the upper bound is attained if and only if
\begin{equation}
\left\langle e^{i\alpha}\right\rangle_P=0.
\label{eq:zero-first-moment}
\end{equation}
Because
\(\langle e^{-i\alpha}\rangle_P=
\langle e^{i\alpha}\rangle_P^*\), this condition is equivalent to
\(\bm m=0\). The population-dependent source then disappears from the
coherence equation, while the coherence return term disappears from the
population equations. The population and coherence sectors therefore
decouple at the generator level, rather than only for one stationary
solution.

The optimal distribution is not unique when arbitrarily many phases are
allowed. For example, a uniform phase distribution and symmetric multiphase
sets can also satisfy Eq.~\eqref{eq:zero-first-moment}. The optimal
realization using exactly two distinct phase settings with nonzero weights
is, however, unique. If the phases \(\alpha_1\) and \(\alpha_2\) occur with
probabilities \(w\) and \(1-w\), respectively, the zero-first-moment
condition becomes
\(w e^{i\alpha_1}+(1-w)e^{i\alpha_2}=0\). It can hold only when
\begin{equation}
w=\frac{1}{2},
\qquad
\alpha_2-\alpha_1=\pi\pmod{2\pi}.
\label{eq:minimal-optimum}
\end{equation}
Thus, the balanced \(\pi\)-separated pair is the unique optimal realization
using two phase settings. The pair may be shifted by a common phase or its
two labels may be interchanged without violating the zero-first-moment
condition.

The next subsection converts this rate bound into a stationary cooling-power
bound.

\subsection{From the rate bound to a cooling-power bound}

Define the cold and hot transition currents into the working medium as
\(j_c\equiv\bar\lambda_c p_1^{\rm ss}-\lambda_c p_2^{\rm ss}\) and
\(j_h\equiv\bar\lambda_h p_1^{\rm ss}-\lambda_h p_3^{\rm ss}\). The
stationary forms of Eqs.~\eqref{eq:pdot2} and~\eqref{eq:pdot3} give the common
cycle current \(j_c=J_m=-j_h\equiv j\). After the stationary coherence has been absorbed into
\(k_{\rm eff}\), the stationary population-balance equations are exactly
equivalent to those of a classical three-state network. This is a stationary
mapping and does not imply an exact time-local population generator away from
stationarity. Solving the balance equations together with normalization gives
\begin{equation}
j=\frac{k_{\rm eff}
(\bar\lambda_c\lambda_h-\bar\lambda_h\lambda_c)}
{\lambda_c\lambda_h+\bar\lambda_c\lambda_h+\bar\lambda_h\lambda_c
+k_{\rm eff}(\lambda_c+\lambda_h+2\bar\lambda_c+2\bar\lambda_h)}.
\label{eq:closed-current}
\end{equation}
Equation~\eqref{eq:closed-current} is obtained from the directed spanning
trees of the effective network in Appendix~\ref{app:stationary-network},
where its derivative with respect to \(k_{\rm eff}\) is also evaluated.

\begin{figure*}[t]
\centering
\hspace*{0.04\textwidth}%
\begin{minipage}{0.24\textwidth}
\centering
\begin{overpic}[width=\linewidth]{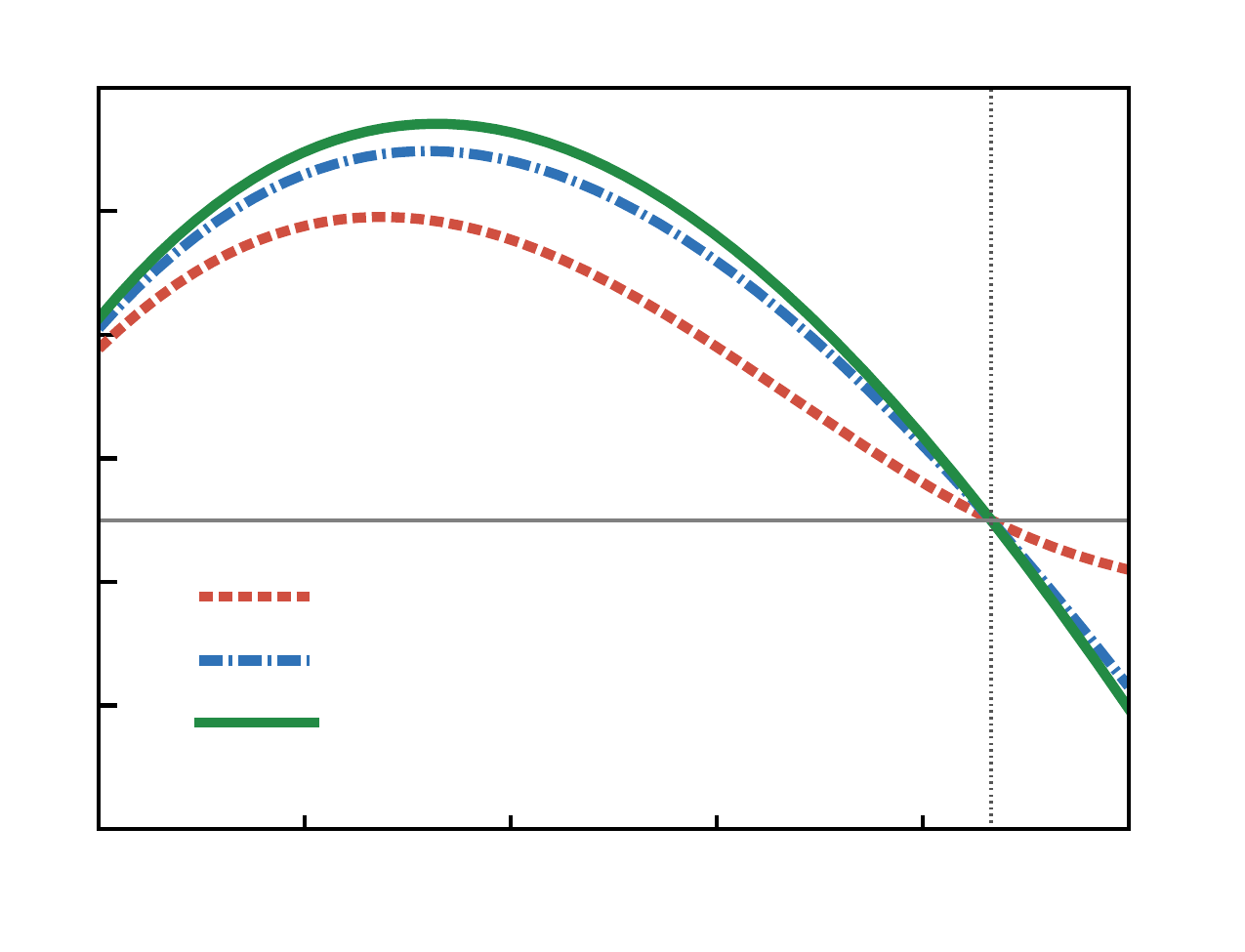}

\put(7,72){(a)}

\put(46,-5){\scalebox{1.15}{$\Delta_{21}$}}
\put(-17,24){\rotatebox{90}{$\dot{Q}_c\ (10^{-3})$}}

\put(1,3){$0.10$}
\put(19,3){$0.25$}
\put(34,3){$0.40$}
\put(51,3){$0.55$}
\put(67,3){$0.70$}
\put(84,3){$0.85$}

\put(-9,9){$-0.5$}
\put(-9,18){$-0.3$}
\put(-9,28){$-0.1$}
\put(-5,38){$0.1$}
\put(-4,48){$0.3$}
\put(-4,58){$0.5$}
\put(-4,67){$0.7$}

\put(27,28){$\delta=0$}
\put(27,22){$\delta=\pi/2$}
\put(27,16){$\delta=\pi$}

\end{overpic}
\end{minipage}%
\hfill
\begin{minipage}{0.24\textwidth}
\centering
\begin{overpic}[width=\linewidth]{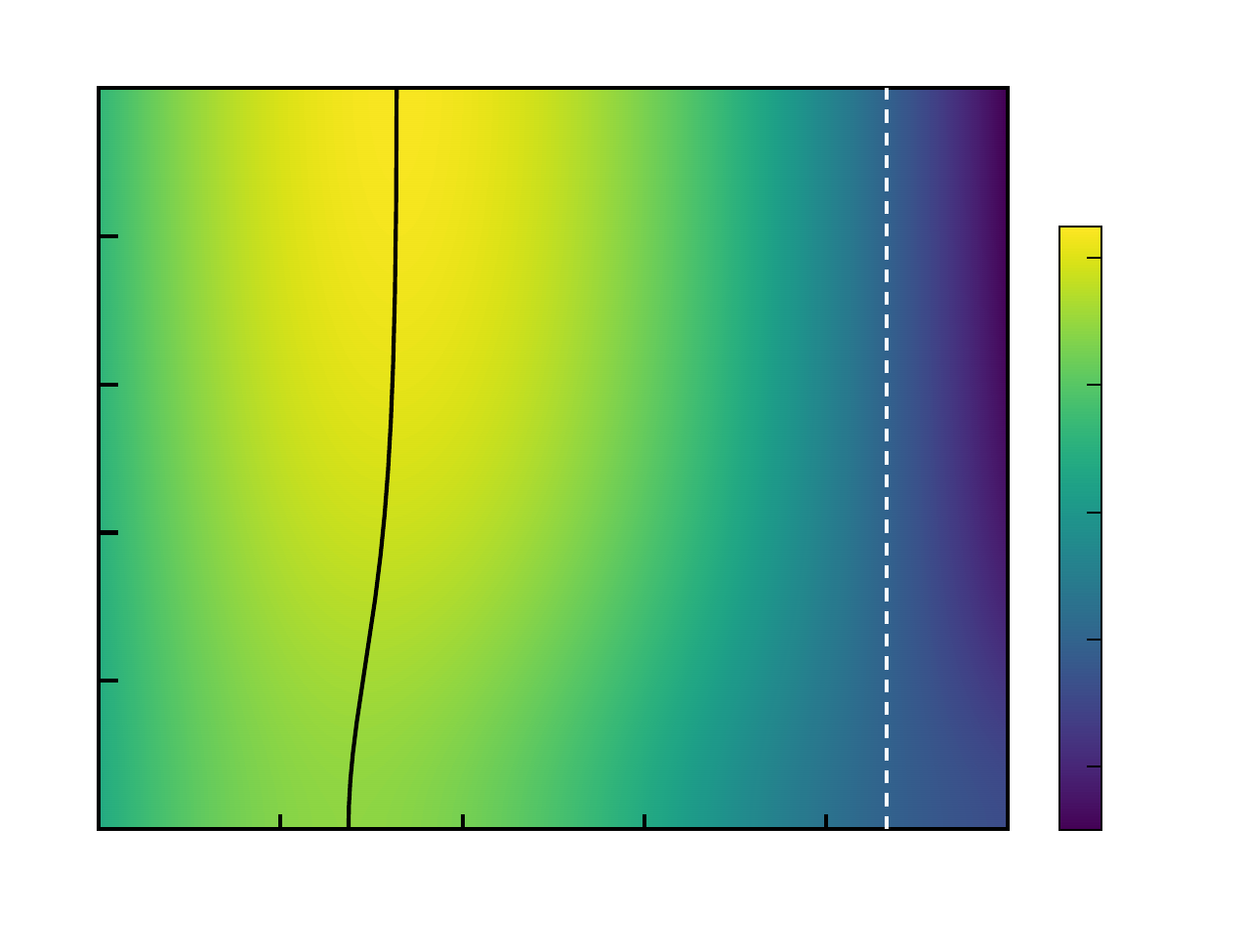}

\put(7,72){(b)}

\put(40,-5){\scalebox{1.15}{$\Delta_{21}$}}
\put(-14,34){\rotatebox{90}{\scalebox{1}{$\delta/\pi$}}}
\put(83,64){$\dot{Q}_c\ (10^{-3})$}

\put(1,3){$0.10$}
\put(16,3){$0.25$}
\put(31,3){$0.40$}
\put(46,3){$0.55$}
\put(62,3){$0.70$}
\put(77,3){$0.85$}

\put(-5,10){$0.0$}
\put(-5,21){$0.2$}
\put(-5,32){$0.4$}
\put(-5,44){$0.6$}
\put(-5,55){$0.8$}
\put(-5,67){$1.0$}

\put(90,14){$-0.2$}
\put(90,24){$0.0$}
\put(90,34){$0.2$}
\put(90,44){$0.4$}
\put(90,54){$0.6$}

\end{overpic}
\end{minipage}%
\hfill
\begin{minipage}{0.24\textwidth}
\centering
\begin{overpic}[width=\linewidth]{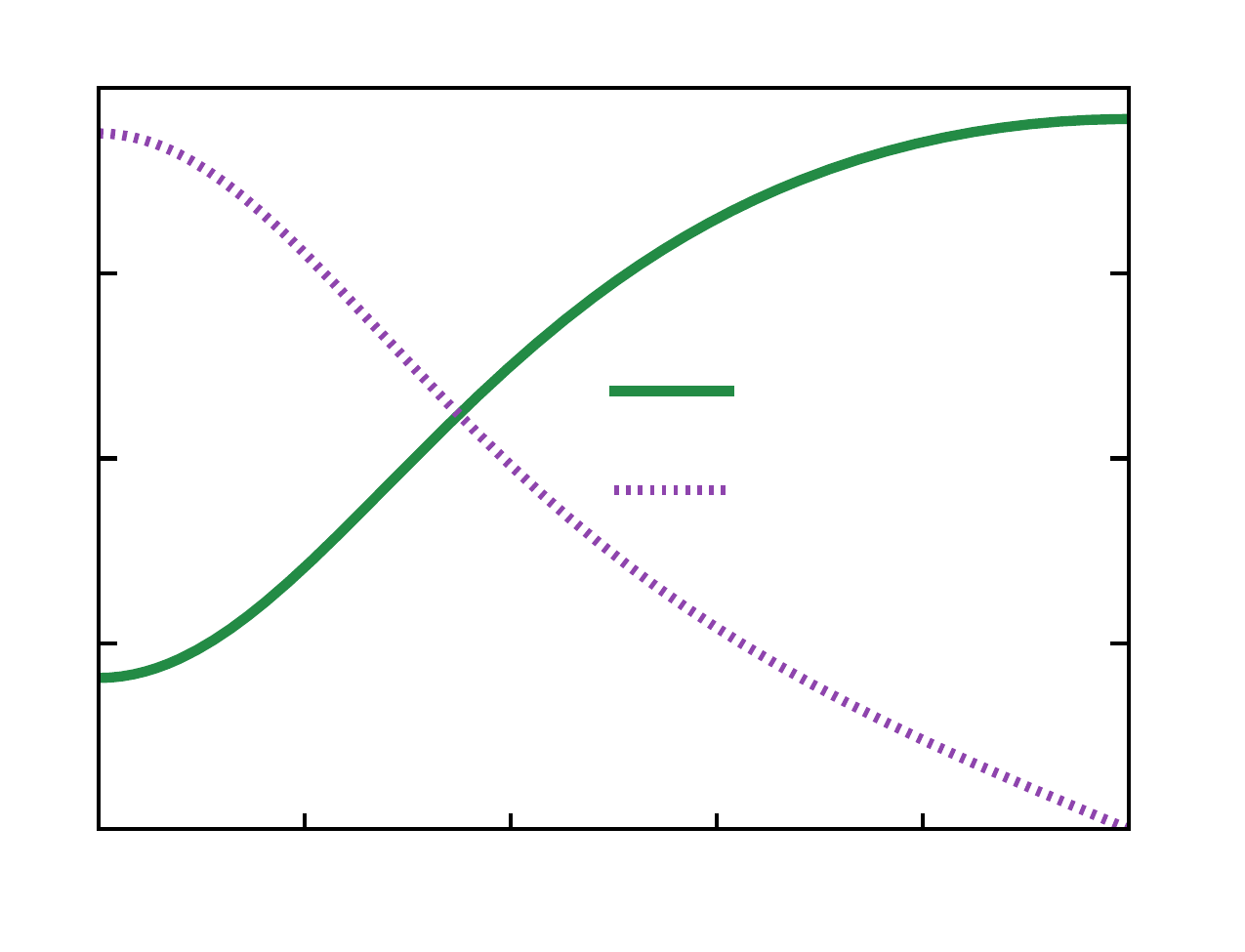}

\put(7,72){(c)}

\put(18,-5){\scalebox{1.15}{phase separation $\delta/\pi$}}
\put(-16,20){\rotatebox{90}{$\dot{Q}_c^{\max}\ (10^{-3})$}}
\put(110,36){\rotatebox{90}{$C_{\ell_1}^{\rm ss}$}}

\put(5,3){$0.0$}
\put(20,3){$0.2$}
\put(37,3){$0.4$}
\put(54,3){$0.6$}
\put(70,3){$0.8$}
\put(86,3){$1.0$}

\put(-7,10){$0.45$}
\put(-7,23){$0.50$}
\put(-7,38){$0.55$}
\put(-7,52){$0.60$}
\put(-7,68){$0.65$}

\put(92,10){$0.000$}
\put(92,23){$0.005$}
\put(92,38){$0.010$}
\put(92,53){$0.015$}
\put(92,67){$0.020$}

\put(59,44){$\dot{Q}_c^{\max}$}
\put(59,35){$C_{\ell_1}^{\rm ss}$}

\end{overpic}
\end{minipage}%
\hspace*{0.04\textwidth}%
\par
\vspace{2mm}
\caption{Cooling consequence of removing coherence feedback.
(a) \(\dot Q_c\) versus \(\Delta_{21}\) for
\(\delta=0,\pi/2,\pi\). The vertical dotted line marks the cooling boundary
\(\Delta_{21}/T_c=\Delta_{31}/T_h\). (b) Cooling landscape in the
\((\Delta_{21},\delta)\) plane; the white dashed line marks
\(\dot Q_c=0\), and the solid line follows the maximum-power ridge. (c)
Cooling power optimized over \(\Delta_{21}\) within the
cooling window, together with the stationary coherence
\(C_{\ell_1}^{\rm ss}\) on the same optimizing trajectory. Parameters are as
in Fig.~\ref{fig:mechanism}.}
\label{fig:cooling}
\end{figure*}

The numerator of Eq.~\eqref{eq:closed-current} is the forward cycle product
minus the backward cycle product. Using local detailed balance, it is
positive precisely in the refrigeration window
\begin{equation}
\frac{\Delta_{21}}{T_c}<\frac{\Delta_{31}}{T_h}.
\label{eq:cooling-window}
\end{equation}
Within this window, \(\partial j/\partial k_{\rm eff}>0\), as established
in Appendix~\ref{app:stationary-network}. Since
\(\dot Q_c=\Delta_{21}j\), combining this monotonicity with
Eq.~\eqref{eq:rate-bound} converts the global rate bound into the fixed-resource
cooling-power bound
\begin{equation}
\dot Q_c[P]\leq\dot Q_c^{\rm opt}
\equiv\dot Q_c(k_{\rm eff}=A_m).
\label{eq:power-bound}
\end{equation}
Every zero-first-moment distribution in the independent-sampling class
reaches this bound, and the balanced \(\pi\)-separated pair uses the smallest
possible number of phase settings.

The cooling curves in Fig.~\ref{fig:cooling}(a) illustrate the consequence
of the rate bound at three phase separations. For every gap inside the cooling window, increasing the phase
separation raises \(k_{\rm eff}\) and therefore raises \(\dot Q_c\). The
three curves cross zero at the same value
\(\Delta_{21}=T_c\Delta_{31}/T_h=0.75\), because the phase protocol changes
the stationary upper-link coefficient but not the detailed-balance condition that
sets the thermodynamic direction of the cycle. Outside the window the current
reverses, and increasing \(k_{\rm eff}\) enhances the reversed cycle rather
than creating refrigeration. At the representative gap
\(\Delta_{21}=0.34\), the change from \(\delta=0\) to \(\delta=\pi\)
increases the cooling power from \(4.86\times10^{-4}\) to
\(6.41\times10^{-4}\), a gain of about \(32.0\%\).

The landscape in Fig.~\ref{fig:cooling}(b) confirms that this improvement
is not confined to an isolated gap. The power increases smoothly toward \(\delta=\pi\) throughout
the cooling region, while the vertical zero-current line remains independent
of \(\delta\). The optimal-power ridge shifts because the upper-state rate
changes the balance between energy extracted per cycle and cycle frequency.

When the gap is optimized separately at each phase separation
[Fig.~\ref{fig:cooling}(c)], the maximum power rises monotonically while
the coherence falls to zero. For \(\delta=0\) and \(\pi\), the optimal gaps are approximately
\(0.306\) and \(0.346\), and the optimized power increases by \(30.7\%\).
Both optima lie well inside the cooling window, so the gain is not caused by
placing the reference protocol near the reversible boundary.

\begin{figure*}[t!]
\centering
\makebox[\textwidth][c]{%
\hspace*{0.04\textwidth}%
\begin{minipage}{0.24\textwidth}
\centering
\begin{overpic}[width=\linewidth]{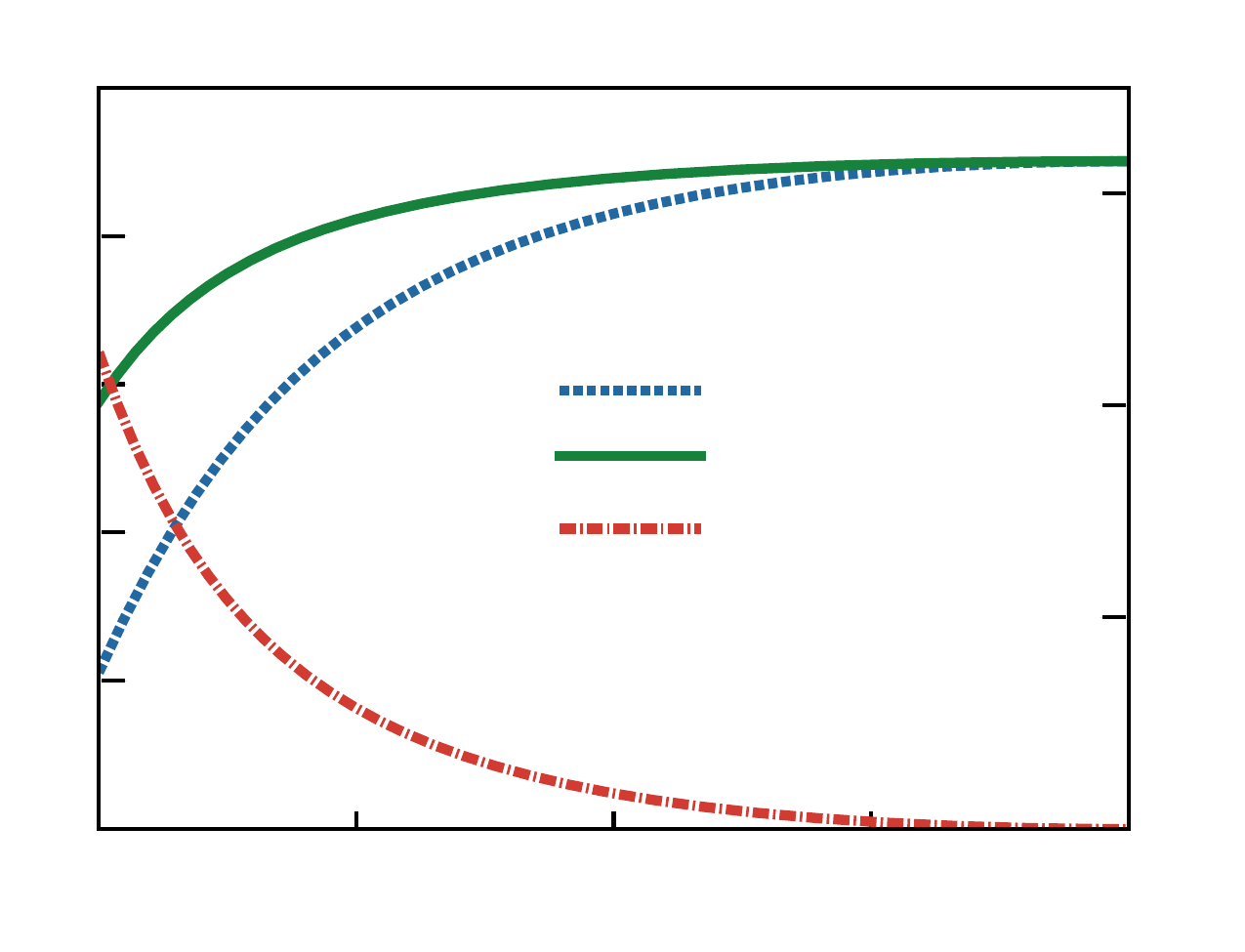}
\put(7,72){(a)}
\put(47,-5){\scalebox{1.15}{$\theta/\pi$}}
\put(-15,20){\rotatebox{90}{$\dot{Q}_c^{\max}\ (10^{-3})$}}
\put(5,3){$0.1$}
\put(24,3){$0.2$}
\put(45,3){$0.3$}
\put(65,3){$0.4$}
\put(85,3){$0.5$}
\put(-5,10){$0.3$}
\put(-5,21){$0.4$}
\put(-5,32){$0.5$}
\put(-5,44){$0.6$}
\put(-5,56){$0.7$}
\put(-5,67){$0.8$}
\put(94,10){$0$}
\put(94,26){$20$}
\put(94,42){$40$}
\put(94,59){$60$}
\put(58,45){$\delta=0$}
\put(58,38){$\delta=\pi$}
\put(58,31){$G(\theta)$}
\end{overpic}
\end{minipage}%
\hspace{0.10\textwidth}%
\begin{minipage}{0.24\textwidth}
\centering
\begin{overpic}[width=\linewidth]{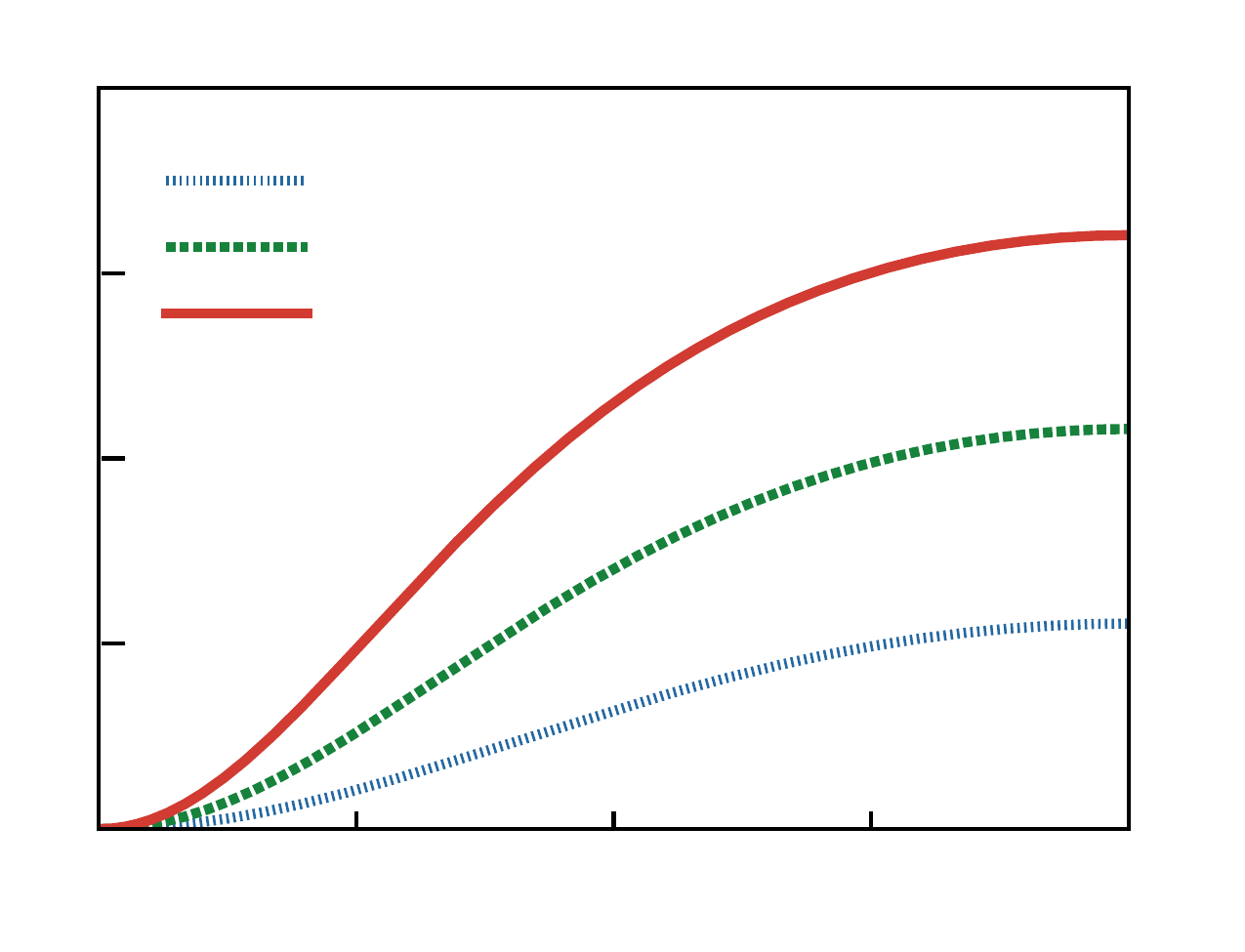}
\put(7,72){(b)}
\put(47,-5){\scalebox{1.15}{$\delta/\pi$}}
\put(-15,20){\rotatebox{90}{$\dot{Q}_c(\delta)/\dot{Q}_c(0)$}}
\put(7,3){$0.00$}
\put(25,3){$0.25$}
\put(45,3){$0.50$}
\put(66,3){$0.75$}
\put(85,3){$1.00$}
\put(-5,10){$1.0$}
\put(-5,24){$1.1$}
\put(-5,38){$1.2$}
\put(-5,53){$1.3$}
\put(-5,67){$1.4$}
\put(26,62){$\gamma_m=0.4$}
\put(26,56){$\gamma_m=0.7$}
\put(26,50){$\gamma_m=1$}
\end{overpic}
\end{minipage}%
\hspace{0.10\textwidth}%
\begin{minipage}{0.24\textwidth}
\centering
\begin{overpic}[width=\linewidth]{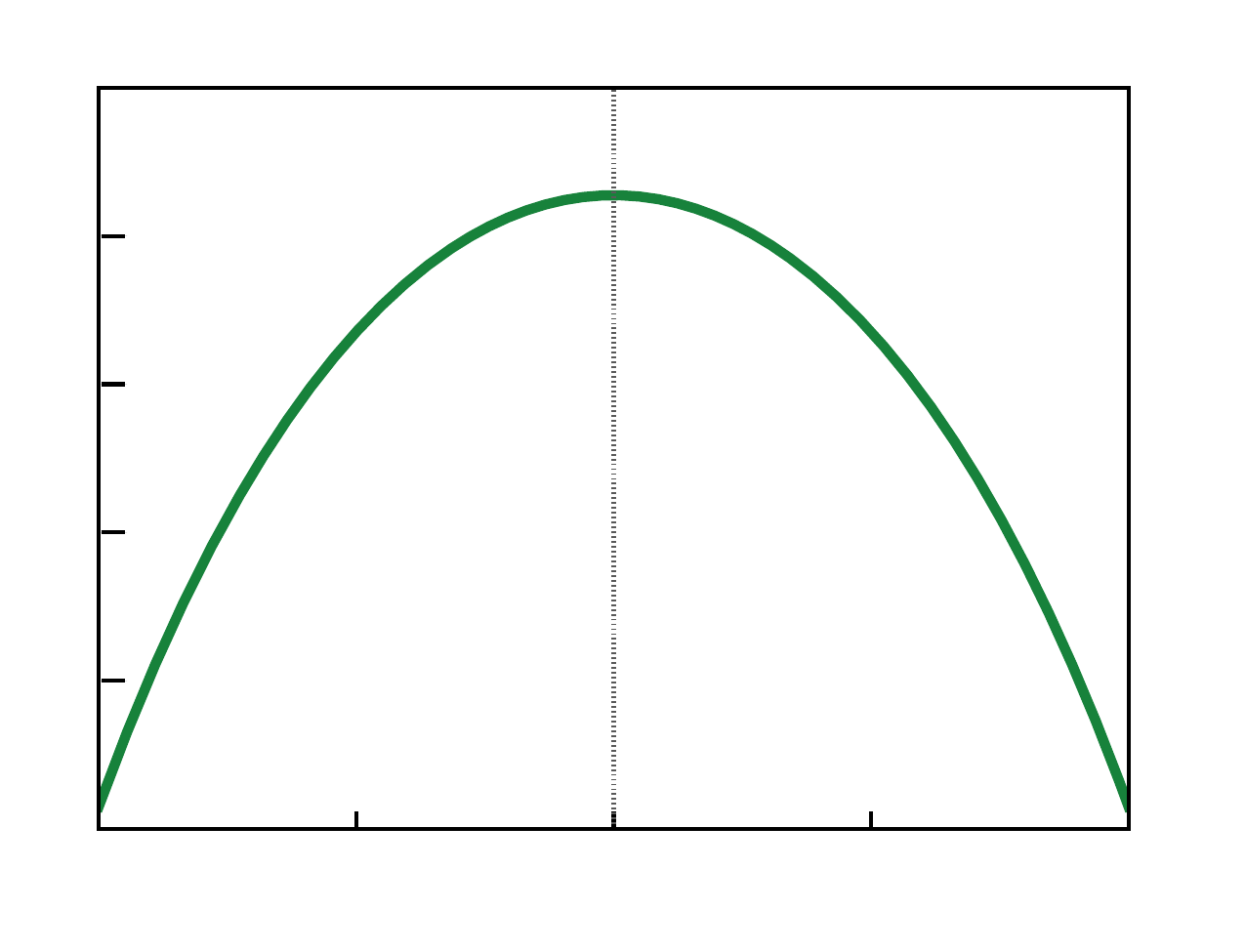}
\put(7,72){(c)}
\put(48,-5){\scalebox{1.15}{$\epsilon$}}
\put(-20,22){\scalebox{1}{\rotatebox{90}{$\dot{Q}_c\ (10^{-3})$}}}
\put(3,3){$-0.50$}
\put(20,3){$-0.25$}
\put(43,3){$0.00$}
\put(64,3){$0.25$}
\put(84,3){$0.50$}
\put(-11,10){$0.620$}
\put(-11,20){$0.625$}
\put(-11,32){$0.630$}
\put(-11,44){$0.635$}
\put(-11,56){$0.640$}
\put(-11,68){$0.645$}
\end{overpic}
\end{minipage}%
\hspace*{0.04\textwidth}%
}
\par
\vspace{2mm}
\caption{Control dependence and robustness. (a) Gap-optimized cooling power
for one fixed phase and for the optimal balanced \(\pi\)-separated pair,
together with the relative gain \(G(\theta)\). The vertical dotted line marks
the representative angle \(\theta=\pi/8\). (b) Fixed-gap power ratio versus
phase separation for \(\gamma_m=0.4,0.7,1.0\), at \(\theta=\pi/8\) and
\(\Delta_{21}=0.34\). (c) Cooling power at \(\delta=\pi\) when the two phases
have probabilities \(w\) and \(1-w\), with duty imbalance
\(\epsilon=2w-1\). Unless stated otherwise, parameters are as in
Fig.~\ref{fig:mechanism}.}
\label{fig:robustness}
\end{figure*}

\section{Energy accounting and working-medium energy ratio}
\label{sec:thermodynamics}

Each forward cycle absorbs the quantum \(\Delta_{21}\) from the cold
reservoir, releases \(\Delta_{31}\) to the hot reservoir, and receives
\(\Delta_{32}=\Delta_{31}-\Delta_{21}\) from the measurement channel.
Multiplying these quanta by the common stationary cycle current gives
\begin{equation}
\dot Q_c=\Delta_{21}j,
\label{eq:Qc}
\end{equation}
\begin{equation}
\dot Q_h=-\Delta_{31}j,
\label{eq:Qh}
\end{equation}
\begin{equation}
\dot E_{\rm meas}=\Tr\!\left\{H_S\mathcal L_P[\rho^{\rm ss}]\right\}
=\Delta_{32}j.
\label{eq:Emeas}
\end{equation}
Equation~\eqref{eq:Emeas} follows by inserting the stationary measurement
population flow into the energy change generated by \(\mathcal L_P\). Using
the gap identity following Eq.~\eqref{eq:HS}, Eqs.~\eqref{eq:Qc}--\eqref{eq:Emeas}
satisfy the exact working-medium first law
\begin{equation}
\dot Q_c+\dot Q_h+\dot E_{\rm meas}=0.
\label{eq:first-law}
\end{equation}
Here \(\dot E_{\rm meas}\) is the energy delivered by the reduced measurement
channel. The reduced map alone does not determine whether a physical
apparatus supplies that energy as work, heat, or a combination of both. Such a
partition is process dependent and can be affected by coherence
~\cite{Su2018}.

For a nonzero cycle current, Eqs.~\eqref{eq:Qc} and~\eqref{eq:Emeas} give the
apparent working-medium energy ratio
\begin{equation}
\varepsilon_{\rm app}=\frac{\dot Q_c}{\dot E_{\rm meas}}
=\frac{\Delta_{21}}{\Delta_{32}}.
\label{eq:energy-ratio}
\end{equation}
At the reversible boundary all three currents vanish, but the right-hand side
remains the ratio of the energy quanta exchanged per completed cycle and is
obtained continuously from the cooling regime. At fixed level spacings,
phase cycling changes only \(j\), so \(\dot Q_c\) and
\(\dot E_{\rm meas}\) increase by the same relative factor. The optimization
proved above is therefore a cooling-power optimization, not an efficiency
optimization.

Only under the additional idealization that \(\dot E_{\rm meas}\) is supplied
entirely as work and all other apparatus costs are negligible does
\(\varepsilon_{\rm app}\) coincide with the conventional work-driven
coefficient of performance. Under that convention,
Eq.~\eqref{eq:cooling-window} implies
\(\varepsilon_{\rm app}<T_c/(T_h-T_c)\), with equality approached only at the
reversible boundary where the power vanishes. A physical detector can incur
additional control, phase-programming, and resetting costs
~\cite{Perna2024,Latune2025,Elouard2026}. If an implementation records outcomes
before discarding them, storage and erasure introduce further
information-processing costs~\cite{Parrondo2015,ReebWolf2014}. A device-level operational
coefficient of performance and the corresponding entropy-production balance
therefore require a microscopic detector model and are not fixed by
Eq.~\eqref{eq:master}.

\section{Control constraints, robustness, and scope}
\label{sec:control}

\subsection{Fixed polar angle and programmable phase}

Measurement-conditioned engines can optimize both the measurement basis and
unitary driving speed~\cite{LiCPL2026}. The phase optimization considered
here instead applies when the system--meter coupling fixes the polar angle
but the relative control phase remains programmable. A fixed
ratio of pulse amplitudes, for example, can set \(\theta\), while the phase of
the control field sets \(\alpha\). This control structure is also explicit in
Eq.~\eqref{eq:phase-covariance}: the same fixed tilted measurement is used at
every event, and only the surrounding relative-phase rotations are changed.
No measurement record is required.

The endpoints clarify the role of the constraint. At \(\theta=\pi/2\), one
has \(A_m=\gamma_m/2\) and \(B_m=C_m=0\), so a single equatorial measurement
already produces the largest direct mixing without a coherence-feedback
path. At \(\theta=0\), no upper-state mixing is generated. Phase cycling is
therefore useful in the constrained interval \(0<\theta<\pi/2\), where the
direct transfer and coherence feedback coexist.

Figure~\ref{fig:robustness}(a) compares the gap-optimized fixed-phase power
\(Q_0^{\max}(\theta)\) with the optimized \(\pi\)-separated power
\(Q_\pi^{\max}(\theta)\), and plots their relative difference
\(G(\theta)\equiv[Q_\pi^{\max}(\theta)-Q_0^{\max}(\theta)]/
Q_0^{\max}(\theta)\). Both absolute powers vanish as \(\theta\to0\), because
\(A_m\propto\sin^2\theta\), even though the relative gain can be large in
that limit. As \(\theta\) increases, the usable direct channel becomes
stronger, while the factor \(\cos\theta\) controlling the coherence loop
shrinks. The two power curves therefore merge and \(G(\theta)\to0\) as
\(\theta\to\pi/2\). The marked value \(\theta=\pi/8\) lies in the regime
where the absolute power remains appreciable and the removed penalty is still
large. The protocol should thus be understood as a correction to a
hardware-constrained tilted basis, not as a method that outperforms a freely
chosen equatorial measurement.

\subsection{Robustness and limits of the reduced model}

The power ratios in Fig.~\ref{fig:robustness}(b) track the full path from
one phase to the balanced \(\pi\)-separated pair. All curves start at unity because they are
normalized to the fixed-phase power at \(\delta=0\), and all increase
smoothly toward \(\delta=\pi\). The absence of an intermediate optimum is the
power-level manifestation of the rate monotonicity proved in
Appendix~\ref{app:two-phase}. A larger \(\gamma_m\) strengthens both the bare mixing
and the coherence loop, so removing the latter produces a larger relative
benefit over the range shown. Crucially, the protocol does not require fine
tuning to an isolated phase separation: a broad neighborhood of the
\(\pi\)-separated point already recovers most of the available power.

The local robustness follows analytically. If
\(\delta=\pi-\varepsilon_\delta\), then
\(c_\delta^2=\sin^2(\varepsilon_\delta/2)\), so
Eq.~\eqref{eq:keff} gives
\(A_m-k_{\rm eff}=O(\varepsilon_\delta^2)\). For unequal probabilities
\(w\) and \(1-w\) of the two \(\pi\)-separated phases, define the duty
imbalance \(\epsilon\equiv2w-1\). The stationary coherence equations are
solved for unequal phase weights in Appendix~\ref{app:two-phase}; their
substitution into the population current gives
\begin{equation}
k_{\rm eff}^{(\epsilon)}(\pi)
=A_m-\frac{B_mC_m\epsilon^2}
{\Gamma_I^\pi+\Delta_{32}^2/\Gamma_R^\pi}.
\label{eq:duty-rate}
\end{equation}
Here \(\Gamma_R^\pi\equiv\gamma_\varphi+\gamma_m\) and
\(\Gamma_I^\pi\equiv\gamma_\varphi+\gamma_m\cos^2\theta\). The residual
coherence is linear in \(\epsilon\), and its coupling back to the populations
is also linear in \(\epsilon\), so the current penalty begins at
\(O(\epsilon^2)\).

The resulting dependence on duty imbalance is shown in
Fig.~\ref{fig:robustness}(c). The maximum is
centered at \(\epsilon=0\), where the two phase weights are equal, and the
curve is symmetric under \(\epsilon\to-\epsilon\), which merely exchanges
the labels of the two phase settings. Its locally flat top is the numerical
signature of the quadratic correction in Eq.~\eqref{eq:duty-rate}. Thus both
the phase-separation error in panel (b) and the probability imbalance in
panel (c) are first-order suppressed around the optimal protocol.

After the stationary coherence is eliminated, the stationary population
balance is represented exactly by a classical three-state network with
upper-link coefficient \(k_{\rm eff}\). Coherence is nevertheless the
microscopic origin of the reduced coefficient for a fixed tilted measurement.
The result identifies and removes a quantum-coherent penalty; it does not
claim a uniquely quantum stationary advantage.

Finally, the theorem assumes independent phase draws and instantaneous Poisson
measurement events. A deterministic alternating phase cycle or a correlated
phase process generally produces time-dependent or history-dependent dynamics
and is not covered by Eq.~\eqref{eq:LP}. Finite measurement duration can also
modify the effective dynamics and external resource cost~\cite{YiKim2026}.
These extensions require a separate implementation-level analysis.

\section{Conclusion}

We formulated randomized open-loop phase cycling over independent phase
distributions \(P(\alpha)\), at fixed event rate, polar angle, bath parameters,
and level spacings. A tilted nonselective measurement combines
phase-independent population mixing with a phase-tagged coherence source.
Stationary elimination gives
\(0\leq k_{\rm eff}[P]\leq A_m\), and the upper bound is attained precisely
when \(\langle e^{i\alpha}\rangle_P=0\).

Two equally weighted antipodal phases provide the minimal realization. They
cancel the population-to-coherence source and its return term at the generator
level while preserving the direct mixing coefficient \(A_m\). Uniform and
symmetric multiphase distributions can also be optimal, but the antipodal pair
is the unique optimum using exactly two nonzero phase weights, up to a common
shift and relabeling.

The cold current is monotone in \(k_{\rm eff}\) throughout the refrigeration
window, so the same condition maximizes stationary power within the
independent-sampling reduced-map class. The representative fixed-gap and
gap-optimized gains are about \(32\%\) and \(31\%\), respectively. Phase
cycling changes the cycle current but not the energy quanta per cycle, leaving
\(\varepsilon_{\rm app}=\Delta_{21}/\Delta_{32}\) unchanged. It therefore
recovers power suppressed by a quantum-coherent feedback path without
establishing a uniquely quantum stationary advantage or a device-level
efficiency gain. The latter requires a microscopic account of the detector
and phase-control costs.


\begin{acknowledgments}
This work was supported by the National Natural Science Foundation of China
(Grant No.~12465009). J.W. additionally acknowledges support from the Major
Program of Jiangxi Provincial Natural Science Foundation, China (Grant
No.~20224ACB201007).
\end{acknowledgments}

\appendix
\section{Detailed derivation of the upper-state coherence equation}
\label{app:bath}

This appendix derives Eq.~\eqref{eq:bath-coherence} directly from the
Hamiltonian and the two local reservoir generators defined in
Eqs.~\eqref{eq:HS}--\eqref{eq:Lh}. We evaluate their
\((2,3)\) matrix elements separately.

Because \(H_S\) is diagonal in the energy basis,
\begin{equation}
-i\bra{2}[H_S,\rho]\ket{3}
=
-i(E_2-E_3)\rho_{23}
=
i\Delta_{32}\rho_{23}.
\label{app:eq:H23}
\end{equation}

For the cold reservoir, the downward jump operator is
\(A_c=\ket{1}\bra{2}\). Its jump term has no
\((2,3)\) matrix element, while
\(A_c^\dagger A_c=\ket{2}\bra{2}\) gives
\(-\rho_{23}/2\) through the anticommutator. The upward operator
\(A_c^\dagger=\ket{2}\bra{1}\) gives no contribution because its
anticommutator projector is \(\ket{1}\bra{1}\). Thus,
\begin{equation}
\begin{aligned}
\bra{2}\mathcal D[\ket{1}\bra{2}]\rho\ket{3}
&=-\frac12\rho_{23},\\
\bra{2}\mathcal D[\ket{2}\bra{1}]\rho\ket{3}
&=0,
\end{aligned}
\qquad
\bra{2}\mathcal L_c[\rho]\ket{3}
=
-\frac{\lambda_c}{2}\rho_{23}.
\label{app:eq:cold23}
\end{equation}

Similarly, for the hot reservoir,
\(A_h=\ket{1}\bra{3}\) gives the damping contribution
\(-\rho_{23}/2\), whereas the upward operator
\(A_h^\dagger=\ket{3}\bra{1}\) gives zero. Hence,
\begin{equation}
\begin{aligned}
\bra{2}\mathcal D[\ket{1}\bra{3}]\rho\ket{3}
&=-\frac12\rho_{23},\\
\bra{2}\mathcal D[\ket{3}\bra{1}]\rho\ket{3}
&=0,
\end{aligned}
\qquad
\bra{2}\mathcal L_h[\rho]\ket{3}
=
-\frac{\lambda_h}{2}\rho_{23}.
\label{app:eq:hot23}
\end{equation}

Combining Eqs.~\eqref{app:eq:H23}--\eqref{app:eq:hot23}
gives
\begin{equation}
\left.\dot\rho_{23}\right|_{H_S+c+h}
=
\left(
i\Delta_{32}
-\frac{\lambda_c+\lambda_h}{2}
\right)\rho_{23},
\label{app:eq:bath-result}
\end{equation}
which reproduces Eq.~\eqref{eq:bath-coherence} in the main text.

Equation~\eqref{app:eq:bath-result} is homogeneous in
\(\rho_{23}\). The Hamiltonian rotates an existing coherence at frequency
\(\Delta_{32}\), while the two downward reservoir channels damp it at the
total rate \((\lambda_c+\lambda_h)/2\). No term proportional to
\(p_2\), \(p_3\), or \(p_2-p_3\) appears. The independent local
reservoirs therefore cannot generate upper-state coherence from a
diagonal upper-state density matrix. The population-dependent coherence
source discussed in the main text originates entirely from the tilted
measurement map.

\section{Fixed-phase nonselective measurement map}
\label{app:fixed-map}

We derive the state change produced by one measurement event with fixed phase
\(\alpha\), and then convert that event map into the continuous-time Poisson
generator. In the ordered upper-state basis \(\{\ket{2},\ket{3}\}\), write
\begin{align*}
\rho_u&=
\begin{pmatrix}
p_2&\rho_{23}\\
\rho_{32}&p_3
\end{pmatrix}
=
\begin{pmatrix}
p_2&R+iI\\
R-iI&p_3
\end{pmatrix},\\
\Delta p&=p_2-p_3,
\end{align*}
and define the coherence quadrature selected by the phase,
\begin{equation}
U_\alpha\equiv\operatorname{Re}(e^{i\alpha}\rho_{23})
=R\cos\alpha-I\sin\alpha.
\label{app:eq:Ualpha}
\end{equation}
Let \(c=\cos(\theta/2)\) and \(s=\sin(\theta/2)\). The two projectors in the
upper-state subspace are
\begin{align*}
\Pi_+(\alpha)&=
\begin{pmatrix}
c^2&cs e^{-i\alpha}\\
cs e^{i\alpha}&s^2
\end{pmatrix},\\
\Pi_-(\alpha)&=
\begin{pmatrix}
s^2&-cs e^{-i\alpha}\\
-cs e^{i\alpha}&c^2
\end{pmatrix}.
\end{align*}
The discarded-outcome probabilities are
\begin{align*}
q_+&=c^2p_2+s^2p_3+2csU_\alpha,\\
q_-&=s^2p_2+c^2p_3-2csU_\alpha.
\end{align*}
Since \(\Pi_s\rho_u\Pi_s=q_s\Pi_s\), the postmeasurement upper block is
\(q_+\Pi_+(\alpha)+q_-\Pi_-(\alpha)\). Using
\(2cs=\sin\theta\), \(c^2-s^2=\cos\theta\), and
\(q_+-q_-=\cos\theta\,\Delta p+2\sin\theta\,U_\alpha\), direct matrix
multiplication gives
\begin{subequations}\label{app:eq:single-map}
\begin{align}
\bra{2}\mathcal M_{\theta,\alpha}[\rho]\ket{2}
&=p_2-\frac12\sin^2\theta\,\Delta p
+\sin\theta\cos\theta\,U_\alpha,
\label{app:eq:map-p2}\\
\bra{3}\mathcal M_{\theta,\alpha}[\rho]\ket{3}
&=p_3+\frac12\sin^2\theta\,\Delta p
-\sin\theta\cos\theta\,U_\alpha,
\label{app:eq:map-p3}\\
\bra{2}\mathcal M_{\theta,\alpha}[\rho]\ket{3}
&=e^{-i\alpha}\left[
\sin^2\theta\,U_\alpha
+\frac12\sin\theta\cos\theta\,\Delta p
\right].
\label{app:eq:map-rho23}
\end{align}
\end{subequations}
The \((3,2)\) element is the complex conjugate of
Eq.~\eqref{app:eq:map-rho23}. Equations~\eqref{app:eq:map-p2} and
\eqref{app:eq:map-p3} separate phase-independent population mixing from the
conversion of preexisting coherence into a population change. Conversely, the
second term in Eq.~\eqref{app:eq:map-rho23} converts the population imbalance
into a coherence directed by \(e^{-i\alpha}\).

The special cases in the main text follow immediately. With pre-event state
\(\ket{2}\bra{2}\), the probability to find population in \(\ket{3}\) after
the event is \(\sin^2\theta/2\); the same value holds for the reverse
transition. If the upper block is diagonal, \(U_\alpha=0\), and
Eq.~\eqref{app:eq:map-rho23} reduces to
Eq.~\eqref{2m3}.

\section{Derivation of the random-phase measurement generator}
\label{app:random-phase-generator}

We derive the ensemble-averaged generator associated with the random-phase
measurement protocol introduced in the main text. Immediately before the
\(n\)th measurement event, the phase \(\alpha_n\) is selected independently
from the probability density \(P(\alpha)\), with
\(\int_0^{2\pi}P(\alpha)\,d\alpha=1\). The random variables
\(\{\alpha_n\}\) are independent and identically distributed. Their selection
is independent of the system state, all earlier phase choices, and all
measurement outcomes. The protocol is therefore open loop and contains no
measurement-based feedback.

For a specified realization of the event times and phase choices, let
\(\mathcal H_t=\{(t_n,\alpha_n):t_n\leq t\}\) denote the complete classical
history up to time \(t\). Conditioned on this history, the event at time
\(t_n\) updates the state according to
\begin{equation}
\rho_{\mathcal H_t}(t_n^+)
=
\mathcal M_{\theta,\alpha_n}
\left[
\rho_{\mathcal H_t}(t_n^-)
\right],
\label{app:eq:conditional-update}
\end{equation}
where \(t_n^-\) and \(t_n^+\) denote the instants immediately before and
after the event, respectively.

For each fixed phase \(\alpha\), the map
\(\mathcal M_{\theta,\alpha}\) is nonselective and can be written as
\(\mathcal M_{\theta,\alpha}[\rho]
=\sum_r\Pi_r(\theta,\alpha)\rho\Pi_r(\theta,\alpha)\), where \(r\) labels
the possible measurement outcomes. Because the outcome is discarded, this
map has already performed the quantum average over the measurement outcomes.
The remaining randomness is classical and arises from the
Poisson-distributed event times and the independently selected phases.

The unconditional density operator is obtained by averaging the conditional
state over all possible classical histories:
\begin{equation}
\bar{\rho}(t)
=
\mathbb E_{\mathcal H_t}
\left[
\rho_{\mathcal H_t}(t)
\right].
\label{app:eq:unconditional-state}
\end{equation}

The measurement events form a Poisson process with rate \(\gamma_m\). During
a sufficiently short interval \([t,t+dt]\), no event occurs with probability
\(1-\gamma_m dt+\mathcal O(dt^2)\), exactly one event occurs with probability
\(\gamma_m dt+\mathcal O(dt^2)\), and the probability of two or more events
is \(\mathcal O(dt^2)\). Conditioned on the occurrence of one event, the
probability that the selected phase lies in
\([\alpha,\alpha+d\alpha]\) is \(P(\alpha)d\alpha\). Hence, the joint
probability that one event occurs during \(dt\) with phase in this interval
is \(\gamma_m dt\,P(\alpha)d\alpha+\mathcal O(dt^2)\).

If no event occurs, the measurement process leaves the state unchanged. If
one event occurs with phase \(\alpha\), the state is transformed by
\(\mathcal M_{\theta,\alpha}\). Averaging these alternatives gives
\begin{equation}
\begin{aligned}
\bar{\rho}(t+dt)
={}&
\left(1-\gamma_m dt\right)\bar{\rho}(t)
\\
&+
\gamma_m dt
\int_0^{2\pi}
d\alpha\,P(\alpha)
\mathcal M_{\theta,\alpha}
\left[
\bar{\rho}(t)
\right]
+
\mathcal O(dt^2).
\end{aligned}
\label{app:eq:short-time-average}
\end{equation}
Here we have used the linearity of
\(\mathcal M_{\theta,\alpha}\) and the independence of the newly selected
phase from the system state and the earlier classical history.

Subtracting \(\bar{\rho}(t)\) from both sides of
Eq.~\eqref{app:eq:short-time-average}, dividing by \(dt\), and taking the
limit \(dt\to0\), we obtain
\begin{equation}
\left.
\frac{d\bar{\rho}}{dt}
\right|_{P}
=
\gamma_m
\left[
\int_0^{2\pi}
d\alpha\,P(\alpha)
\mathcal M_{\theta,\alpha}[\bar{\rho}]
-
\bar{\rho}
\right].
\label{app:eq:random-phase-master-equation}
\end{equation}
The corresponding random-phase measurement generator is therefore
\begin{equation}
\mathcal L_P[\rho]
=
\gamma_m
\left[
\int_0^{2\pi}
d\alpha\,P(\alpha)
\mathcal M_{\theta,\alpha}[\rho]
-
\rho
\right].
\label{app:eq:random-phase-generator}
\end{equation}

Equation~\eqref{app:eq:random-phase-generator} has the standard structure of
a Poissonian jump generator. The integral
\(\int_0^{2\pi}d\alpha\,P(\alpha)
\mathcal M_{\theta,\alpha}[\rho]\) is the post-event state averaged over
the externally selected phase. Subtracting \(\rho\) gives the average state
change produced by one event, and multiplication by \(\gamma_m\) converts
this single-event change into a change per unit time.

Three distinct averages enter this construction. Each nonselective map
\(\mathcal M_{\theta,\alpha}\) performs the quantum average over the
discarded measurement outcomes. The Poisson statistics produce a classical
average over the random event occurrences, while the integral over
\(P(\alpha)\) produces a classical average over the phase selected at each
event. The two classical sources of randomness are jointly included in the
average over the complete marked Poisson history \(\mathcal H_t\).

As a consistency check, when the phase is fixed at \(\alpha_0\), the
distribution is \(P(\alpha)=\delta_{2\pi}(\alpha-\alpha_0)\), and
Eq.~\eqref{app:eq:random-phase-generator} reduces to
\(\mathcal L_{\alpha_0}[\rho]
=\gamma_m\{\mathcal M_{\theta,\alpha_0}[\rho]-\rho\}\). For the uniform
distribution \(P(\alpha)=1/(2\pi)\), the same equation contains the uniform
phase average of \(\mathcal M_{\theta,\alpha}\).

If the Hamiltonian and reservoir contributions are described by
\(\mathcal L_0\), the complete unconditional master equation is
\(d\rho/dt=\mathcal L_0[\rho]+\mathcal L_P[\rho]\). In the main text and in
the remainder of the paper, the overbar is omitted, and \(\rho\) denotes the
unconditional density operator.

\section{Balanced two-phase average, stationary elimination, and control errors}
\label{app:two-phase}

For fixed \(\alpha\), Poisson events generate
\(\mathcal L_{m,\alpha}=\gamma_m(\mathcal M_{\theta,\alpha}-\mathcal I)\).
Using \(A_m=\gamma_m\sin^2\theta/2\),
\(B_m=\gamma_m\sin\theta\cos\theta\), and \(C_m=B_m/2\),
Eqs.~\eqref{app:eq:single-map} give
\begin{subequations}\label{app:eq:fixed-phase-equations}
\begin{align}
\left.\dot p_2\right|_\alpha
&=-A_m\Delta p+B_mU_\alpha,
\label{app:eq:fixed-p2}\\
\left.\dot p_3\right|_\alpha
&=+A_m\Delta p-B_mU_\alpha,
\label{app:eq:fixed-p3}\\
\left.\dot R\right|_\alpha
&=\gamma_m[\sin^2\theta\cos\alpha\,U_\alpha-R]
+C_m\cos\alpha\,\Delta p,
\label{app:eq:fixed-R}\\
\left.\dot I\right|_\alpha
&=-\gamma_m[\sin^2\theta\sin\alpha\,U_\alpha+I]
-C_m\sin\alpha\,\Delta p.
\label{app:eq:fixed-I}
\end{align}
\end{subequations}
The terms \(\mp A_m\Delta p\) form the direct mixing channel. The terms
\(\pm B_mU_\alpha\) describe feedback of preexisting coherence onto the
populations. The vector
\(C_m(\cos\alpha,-\sin\alpha)^T\Delta p\) is the phase-tagged coherence
source. These fixed-phase equations are the starting point for the following
two appendices.

For \(\alpha_\pm=\pm\delta/2\), define
\(c_\delta=\cos(\delta/2)\) and \(s_\delta=\sin(\delta/2)\). The selected
quadratures are \(U_+=c_\delta R-s_\delta I\) and
\(U_-=c_\delta R+s_\delta I\). The required equal-weight averages are
\begin{align*}
\frac{U_++U_-}{2}&=c_\delta R,\\
\frac{\cos\alpha_+U_++\cos\alpha_-U_-}{2}&=c_\delta^2R,\\
\frac{\sin\alpha_+U_++\sin\alpha_-U_-}{2}&=-s_\delta^2I,\\
\frac{\cos\alpha_++\cos\alpha_-}{2}&=c_\delta,\\
\frac{\sin\alpha_++\sin\alpha_-}{2}&=0.
\end{align*}
Averaging Eqs.~\eqref{app:eq:fixed-phase-equations} therefore gives
\begin{align*}
\left.\dot p_2\right|_{\rm pc}
&=-A_m\Delta p+B_mc_\delta R,\\
\left.\dot p_3\right|_{\rm pc}
&=+A_m\Delta p-B_mc_\delta R,\\
\left.\dot R\right|_{\rm pc}
&=-\gamma_m(1-\sin^2\theta\,c_\delta^2)R
+C_mc_\delta\Delta p,\\
\left.\dot I\right|_{\rm pc}
&=-\gamma_m(1-\sin^2\theta\,s_\delta^2)I.
\end{align*}
Adding the Hamiltonian rotation
\((\dot R,\dot I)=(-\Delta_{32}I,\Delta_{32}R)\), the reservoir dephasing
\(-\gamma_\varphi(R,I)\), and the lower-transition population terms gives
Eqs.~\eqref{eq:population-coherence}. The vanishing average of
\(\sin\alpha_\pm\) explains why the balanced pair has no direct source in the
\(I\) equation.

At stationarity, Eq.~\eqref{eq:Idot} gives
\(I^{\rm ss}=\Delta_{32}R^{\rm ss}/\Gamma_I(\delta)\), and
Eq.~\eqref{eq:Rdot} then gives
\begin{align*}
R^{\rm ss}&=\frac{C_mc_\delta}{D_\delta}\Delta p^{\rm ss},\\
D_\delta&=\Gamma_R(\delta)+\frac{\Delta_{32}^2}{\Gamma_I(\delta)}.
\end{align*}
Substitution into \(J_m=A_m\Delta p^{\rm ss}-B_mc_\delta R^{\rm ss}\)
yields Eq.~\eqref{eq:keff}.

To prove monotonicity, set \(x=c_\delta^2\), so
\(s_\delta^2=1-x\). Then
\begin{align*}
\Gamma_R(x)&=\gamma_\varphi+\gamma_m(1-\sin^2\theta\,x),\\
\Gamma_I(x)&=\gamma_\varphi+
\gamma_m(\cos^2\theta+\sin^2\theta\,x).
\end{align*}
Both rates are positive, and
\begin{align*}
D'(x)&=-\gamma_m\sin^2\theta
-\frac{\Delta_{32}^2\gamma_m\sin^2\theta}{\Gamma_I(x)^2}<0,\\
\frac{d}{dx}\left[\frac{x}{D(x)}\right]
&=\frac{D(x)-xD'(x)}{D(x)^2}>0.
\end{align*}
As \(\delta\) increases from zero to \(\pi\), \(x\) decreases from one to
zero. Hence \(\Delta k_{\rm coh}=B_mC_mx/D(x)\) decreases monotonically and
vanishes at \(\delta=\pi\). A separation error
\(\delta=\pi-\varepsilon_\delta\) gives
\(x=\sin^2(\varepsilon_\delta/2)=O(\varepsilon_\delta^2)\), which proves the
quadratic phase-error sensitivity.

For unequal phase probabilities, let the \(+\) and \(-\) settings occur with
probabilities \(w\) and \(1-w\), and define \(\epsilon=2w-1\). The relevant
moments are \(\langle\cos\alpha\rangle=c_\delta\),
\(\langle\sin\alpha\rangle=\epsilon s_\delta\), and
\(\langle\sin\alpha\cos\alpha\rangle=\epsilon c_\delta s_\delta\).
Averaging the fixed-phase equations gives
\begin{align*}
\left.\dot p_2\right|_{\rm pc}
&=-A_m\Delta p+B_m(c_\delta R-\epsilon s_\delta I),\\
\left.\dot p_3\right|_{\rm pc}
&=+A_m\Delta p-B_m(c_\delta R-\epsilon s_\delta I),\\
\left.\dot R\right|_{\rm pc}
&=-\gamma_m(1-\sin^2\theta\,c_\delta^2)R\\
&\quad-\gamma_m\sin^2\theta\,\epsilon c_\delta s_\delta I
+C_mc_\delta\Delta p,\\
\left.\dot I\right|_{\rm pc}
&=-\gamma_m\sin^2\theta\,\epsilon c_\delta s_\delta R\\
&\quad-\gamma_m(1-\sin^2\theta\,s_\delta^2)I
-C_m\epsilon s_\delta\Delta p.
\end{align*}
At \(\delta=\pi\), after adding Hamiltonian and reservoir terms,
\begin{align*}
\dot R&=-\Delta_{32}I-\Gamma_R^\pi R,\\
\dot I&=\Delta_{32}R-\Gamma_I^\pi I-C_m\epsilon\Delta p,
\end{align*}
where \(\Gamma_R^\pi=\gamma_\varphi+\gamma_m\) and
\(\Gamma_I^\pi=\gamma_\varphi+\gamma_m\cos^2\theta\). The stationary
solution is
\begin{align*}
I^{\rm ss}&=-\frac{C_m\epsilon\Delta p^{\rm ss}}
{\Gamma_I^\pi+\Delta_{32}^2/\Gamma_R^\pi},\\
R^{\rm ss}&=-\frac{\Delta_{32}}{\Gamma_R^\pi}I^{\rm ss}.
\end{align*}
The population current is
\(J_m=A_m\Delta p^{\rm ss}+B_m\epsilon I^{\rm ss}\). Substitution gives
Eq.~\eqref{eq:duty-rate}. The residual coherence and its return coefficient
are both linear in \(\epsilon\), so the current loss is quadratic.

\section{Arbitrary phase distributions and the exact rate bound}
\label{app:general-phase}

For normalized \(P(\alpha)\), define
\(m_c=\langle\cos\alpha\rangle_P\),
\(m_s=\langle\sin\alpha\rangle_P\),
\(q_{cc}=\langle\cos^2\alpha\rangle_P\),
\(q_{ss}=\langle\sin^2\alpha\rangle_P\), and
\(q_{cs}=\langle\sin\alpha\cos\alpha\rangle_P\). We use
\(\bm u=(R,I)^T\) and \(\bm m=(m_c,-m_s)^T\). Averaging
Eqs.~\eqref{app:eq:fixed-phase-equations}, then adding Hamiltonian precession
and reservoir dephasing, gives
\begin{subequations}\label{app:eq:general-system}
\begin{align}
\left.\dot p_2\right|_P
&=-A_m\Delta p+B_m\bm m^T\bm u,
\label{app:eq:general-p2}\\
\left.\dot p_3\right|_P
&=+A_m\Delta p-B_m\bm m^T\bm u,
\label{app:eq:general-p3}\\
\dot{\bm u}
&=(-\bm\Gamma_P+\bm\Omega)\bm u+C_m\bm m\Delta p.
\label{app:eq:general-u}
\end{align}
\end{subequations}
Here
\begin{align*}
\bm\Gamma_P
&=\gamma_\varphi\bm 1+\gamma_m
\begin{pmatrix}
1-\sin^2\theta\,q_{cc}&\sin^2\theta\,q_{cs}\\
\sin^2\theta\,q_{cs}&1-\sin^2\theta\,q_{ss}
\end{pmatrix},\\
\bm\Omega
&=\Delta_{32}
\begin{pmatrix}0&-1\\1&0\end{pmatrix}.
\end{align*}
The same first moment \(\bm m\) controls both directions of the feedback
loop: \(C_m\bm m\Delta p\) creates coherence and
\(B_m\bm m^T\bm u\) returns it to the populations.

Introduce
\begin{align*}
\bm Q
&=\begin{pmatrix}q_{cc}&-q_{cs}\\-q_{cs}&q_{ss}\end{pmatrix}\\
&=\left\langle
\begin{pmatrix}\cos\alpha\\-\sin\alpha\end{pmatrix}
\begin{pmatrix}\cos\alpha&-\sin\alpha\end{pmatrix}
\right\rangle_P.
\end{align*}
The matrix \(\bm Q\) is positive semidefinite and has unit trace, so its
eigenvalues lie in \([0,1]\). Since
\(\bm\Gamma_P=\gamma_\varphi\bm 1+
\gamma_m(\bm 1-\sin^2\theta\,\bm Q)\),
\begin{equation}
\bm\Gamma_P\succeq
(\gamma_\varphi+\gamma_m\cos^2\theta)\bm 1>0.
\label{app:eq:Gamma-positive}
\end{equation}
Thus \(\bm\Gamma_P-\bm\Omega\) is invertible. Setting
Eq.~\eqref{app:eq:general-u} to zero gives
\(\bm u^{\rm ss}=C_m(\bm\Gamma_P-\bm\Omega)^{-1}
\bm m\Delta p^{\rm ss}\), and substitution into
Eq.~\eqref{app:eq:general-p2} gives Eq.~\eqref{eq:general-keff}.

To prove the sign of the correction, let
\(\bm y=(\bm\Gamma_P-\bm\Omega)^{-1}\bm m\). Since
\(\bm m=(\bm\Gamma_P-\bm\Omega)\bm y\),
\begin{align*}
\bm m^T(\bm\Gamma_P-\bm\Omega)^{-1}\bm m
&=\bm y^T(\bm\Gamma_P+\bm\Omega)\bm y\\
&=\bm y^T\bm\Gamma_P\bm y\geq0.
\end{align*}
The antisymmetric term vanishes because \(\bm y^T\bm\Omega\bm y=0\). Since
\(\bm\Gamma_P\) is positive definite, the correction is zero if and only if
\(\bm m=0\). This proves the upper bound and the saturation condition.

For the lower bound, define
\(\bm K=\bm\Gamma_P^{-1/2}\bm\Omega\bm\Gamma_P^{-1/2}\) and
\(\bm z=\bm\Gamma_P^{-1/2}\bm m\). The matrix \(\bm K\) is antisymmetric.
The antisymmetric part of \((\bm 1-\bm K)^{-1}\) does not contribute to a
real quadratic form, and therefore
\begin{align*}
\bm m^T(\bm\Gamma_P-\bm\Omega)^{-1}\bm m
&=\bm z^T(\bm 1-\bm K)^{-1}\bm z\\
&=\bm z^T(\bm 1-\bm K^2)^{-1}\bm z\\
&\leq\bm z^T\bm z.
\end{align*}
Because \(\|\bm m\|\leq1\), Eq.~\eqref{app:eq:Gamma-positive} gives
\(\bm z^T\bm z\leq1/(\gamma_m\cos^2\theta)\) when
\(\cos\theta\neq0\). Multiplication by
\(B_mC_m=\gamma_m^2\sin^2\theta\cos^2\theta/2\) bounds the correction by
\(A_m=\gamma_m\sin^2\theta/2\). If \(\cos\theta=0\), then
\(B_mC_m=0\). This completes the proof of Eq.~\eqref{eq:rate-bound}.

For two nonzero phase weights, the zero-first-moment condition is
\(w e^{i\alpha_1}+(1-w)e^{i\alpha_2}=0\). The vectors can cancel only if
their magnitudes are equal, \(w=1/2\), and their directions are opposite,
\(\alpha_2-\alpha_1=\pi\pmod{2\pi}\). This proves
Eq.~\eqref{eq:minimal-optimum}. More complicated zero-first-moment
distributions are also optimal, but none uses fewer than two settings.

\section{Stationary population network and cooling-current monotonicity}
\label{app:stationary-network}

After eliminating the stationary coherence through
Eq.~\eqref{eq:general-keff}, the populations obey
\begin{align*}
0&=-(\bar\lambda_c+\bar\lambda_h)p_1+\lambda_cp_2+\lambda_hp_3,\\
0&=\bar\lambda_cp_1-\lambda_cp_2-k_{\rm eff}(p_2-p_3),\\
0&=\bar\lambda_hp_1-\lambda_hp_3+k_{\rm eff}(p_2-p_3),
\end{align*}
where \(k_{\rm eff}=k_{\rm eff}[P]\). The exact stationary population
generator is
\begin{equation}
\mathsf W_{\rm eff}=
\begin{pmatrix}
-\bar\lambda_c-\bar\lambda_h&\lambda_c&\lambda_h\\
\bar\lambda_c&-\lambda_c-k_{\rm eff}&k_{\rm eff}\\
\bar\lambda_h&k_{\rm eff}&-\lambda_h-k_{\rm eff}
\end{pmatrix}.
\label{app:eq:Weff}
\end{equation}
The off-diagonal entries are non-negative by Eq.~\eqref{eq:rate-bound}. The
directed spanning-tree weights rooted at states \(1,2,3\) are
\begin{align*}
\tau_1&=\lambda_c\lambda_h+k_{\rm eff}(\lambda_c+\lambda_h),\\
\tau_2&=\bar\lambda_c\lambda_h
+k_{\rm eff}(\bar\lambda_c+\bar\lambda_h),\\
\tau_3&=\bar\lambda_h\lambda_c
+k_{\rm eff}(\bar\lambda_c+\bar\lambda_h).
\end{align*}
With \(\mathcal N=\tau_1+\tau_2+\tau_3\),
\(p_i^{\rm ss}=\tau_i/\mathcal N\), and
\begin{align*}
\mathcal N={}&\lambda_c\lambda_h+
\bar\lambda_c\lambda_h+\bar\lambda_h\lambda_c\\
&+k_{\rm eff}(\lambda_c+\lambda_h+2\bar\lambda_c+2\bar\lambda_h).
\end{align*}
The upper-link current is
\begin{align*}
j&=k_{\rm eff}(p_2^{\rm ss}-p_3^{\rm ss})\\
&=\frac{k_{\rm eff}(\tau_2-\tau_3)}{\mathcal N}\\
&=\frac{k_{\rm eff}
(\bar\lambda_c\lambda_h-\bar\lambda_h\lambda_c)}{\mathcal N},
\end{align*}
which reproduces Eq.~\eqref{eq:closed-current}. The same expression follows
from the cold link and the negative of the hot link.

Let \(A\equiv\bar\lambda_c\lambda_h-
\bar\lambda_h\lambda_c\) and
\(N_0\equiv\lambda_c\lambda_h+\bar\lambda_c\lambda_h+
\bar\lambda_h\lambda_c\). Differentiation gives
\begin{equation}
\frac{\partial j}{\partial k_{\rm eff}}=\frac{A N_0}{\mathcal N^2}.
\label{app:eq:current-derivative}
\end{equation}
Local detailed balance gives
\(A=\lambda_c\lambda_h[e^{-\Delta_{21}/T_c}-e^{-\Delta_{31}/T_h}]\).
Thus \(A>0\) precisely under Eq.~\eqref{eq:cooling-window}, and the derivative
is strictly positive in the refrigeration regime. Together with
Eq.~\eqref{eq:rate-bound}, this proves Eq.~\eqref{eq:power-bound}.

One forward cycle carries \(\Delta_{21}\) into the working medium from the
cold reservoir, \(\Delta_{31}\) out to the hot reservoir, and
\(\Delta_{32}\) into the working medium from the measurement channel. This
gives Eqs.~\eqref{eq:Qc}--\eqref{eq:Emeas} and their sum,
Eq.~\eqref{eq:first-law}.


\end{document}